\documentclass[12pt]{article}
\usepackage{amsmath, amssymb}
\usepackage{xcolor}
\usepackage{multirow}
\usepackage{algorithm}
\usepackage{algpseudocode}
\usepackage{caption}
\usepackage{graphicx}
\usepackage{subcaption}
\usepackage{mathtools}
\usepackage{enumitem}
\usepackage{tikz}
\usepackage{wrapfig}
\usepackage{caption}
\usepackage{mathtools}
\usepackage{fancybox}
\usepackage{hyperref}
\usetikzlibrary{arrows.meta, tikzmark}

\usepackage{tikz,tikz-3dplot}
\usepackage{pgfplots}
\usepackage{ifthen}
\usepackage{boxedminipage}
\usepackage{makecell}
\usepackage{titlesec}
\titleformat{\section}{\bfseries\large}{\thesection.}{0.5em}{}
\titleformat{\subsection}{\itshape}{\thesubsection.}{0.5em}{}
\usepackage{caption}
\usepackage{array}
\usepackage{lipsum}
\usepackage{float}
\usepackage{booktabs}
\usepackage{stfloats}

\usepackage[margin=1in]{geometry}
\usepackage[authoryear,longnamesfirst]{natbib} 
\usepackage{hyperref}
\usepackage{url}

\begin{document}

\title{A Highly Scalable TDMA for GPUs and Its Application to Flow Solver Optimization}

\author{
  Seungchan Kim$^{1}$ \and
  Jihoo Kim$^{1}$ \and
  Sanghyun Ha$^{2}$ \and
  Donghyun You$^{1}$\thanks{Corresponding author. Email: \href{mailto:dhyou@postech.ac.kr}{dhyou@postech.ac.kr}. ORCID: 0000-0003-2470-5411}
}

\date{
{\small
$^{1}$ Department of Mechanical Engineering, Pohang University of Science and Technology, 77 Cheongam-ro, Nam-gu, Pohang 37673, Gyeongbuk, Republic of Korea\\
$^{2}$ Samsung Electronics Co., Ltd., 129 Samsung-ro, Yeongtong-gu, Suwon-si 16677, Gyeonggi-do, Republic of Korea
}
}

\maketitle

\begin{abstract}
A tridiagonal matrix algorithm (TDMA), \textbf{Pipelined-TDMA}, is developed for multi-GPU systems to resolve the scalability bottlenecks caused by the sequential structure of conventional divide-and-conquer TDMA. 
The proposed method pipelines multiple tridiagonal systems, overlapping communication with computation and executing {GPU} kernels concurrently to hide non-scalable stages behind scalable compute stages.
To maximize performance, the batch size is optimized to strike a balance between GPU occupancy and pipeline efficiency: larger batches improve throughput for solving tridiagonal systems, while excessively large batches reduce pipeline utilization. 
Performance evaluations on up to 64 NVIDIA A100 GPUs using a one-dimensional (1D) slab-type domain decomposition confirm that, except for the terminal phase of the pipeline, the proposed method successfully hides most of the non-scalable execution time—specifically inter-GPU communication and low-occupancy computation.
The solver achieves ideal weak scaling up to 64 GPUs with one billion grid cells per GPU and reaches 74.7\% of ideal performance in strong scaling tests for a 4-billion-cell problem, relative to a 4-GPU baseline. 
The optimized TDMA is integrated into a ADI-based fractional-step method to remove the scalability bottleneck in the Poisson solver of the flow solver by \citet{ha2021multi}. In a 9-billion-cell simulation on 64 GPUs, the TDMA component in the Poisson solver is accelerated by 4.37×, contributing to a 1.31× overall speedup of the complete flow solver.
\\

\end{abstract}

\maketitle

\section{Introduction}
\label{sec:sample1}
Tridiagonal matrix (TDM) systems serve as a key component in numerical simulations of fluid flow, because they enable the use of implicit schemes for stable solutions while maintaining excellent computational efficiency. 
In particular, the Thomas algorithm~\citep{thomas1949elliptic}, a widely used sequential tridiagonal matrix algorithm (TDMA), solves a TDM system of size $n$ with only $\mathcal{O}(n)$ operations, making it highly efficient.
The efficiency of TDMA is especially valuable in semi-implicit fractional-step methods with second-order spatial discretization in finite-difference methods (FDM) for incompressible flow simulations, as solving multiple TDM systems constitutes the majority of the total computational cost. 
Since the pioneering work of \citet{kim1985application}, such methods have been widely adopted to simulate incompressible flows~\citep{le1990improvement, hahn2002direct, abe2001direct}.
These approaches consist of two main steps—solving the momentum and Poisson equations—both of which involve TDM systems that dominate computational cost. 
In the momentum equations, implicit schemes such as the alternating direction implicit (ADI) method introduce TDM systems through implicit treatment of viscous terms. Similarly, FFT-based Poisson solvers require TDM systems along specific directions and these TDM systems contribute significantly to the total execution time.
For this reason, optimizing TDMA is critical for achieving high-performance incompressible flow simulations.

Despite the efficiency of the Thomas algorithm, a parallel TDMA tailored to massively parallel architectures is crucial for achieving high performance on GPU systems.
There has been extensive research focused on developing parallel TDMAs even before the advent of GPUs~\citep{hockney1965fast, stone1973efficient, hockney1981parallel, wang1981parallel, sun1989parallel, mehrmann1993divide, mattor1995algorithm, polizzi2007spike}.
By leveraging two of these parallel TDMAs—parallel cyclic reduction (PCR)\citep{hockney1981parallel} and cyclic reduction (CR)\citep{hockney1965fast}—\citet{zhang2010fast} proposed a hybrid CR–PCR method, which proves highly effective on a single GPU.
The PCR phase takes advantage of CUDA’s fast on-chip shared memory, minimizing memory access latency and boosting throughput.
Building on this approach, NVIDIA provides a built-in function, \texttt{cusparseDgtsv2}, which solves a TDM system on a single GPU using the hybrid CR–PCR method. 
However, the library does not support solving each TDM system across multiple GPUs.

\begin{figure*}[htbp]
    \centering
    \begin{subfigure}{0.32\textwidth}
        \centering
        \includegraphics[page=1, width=\linewidth]{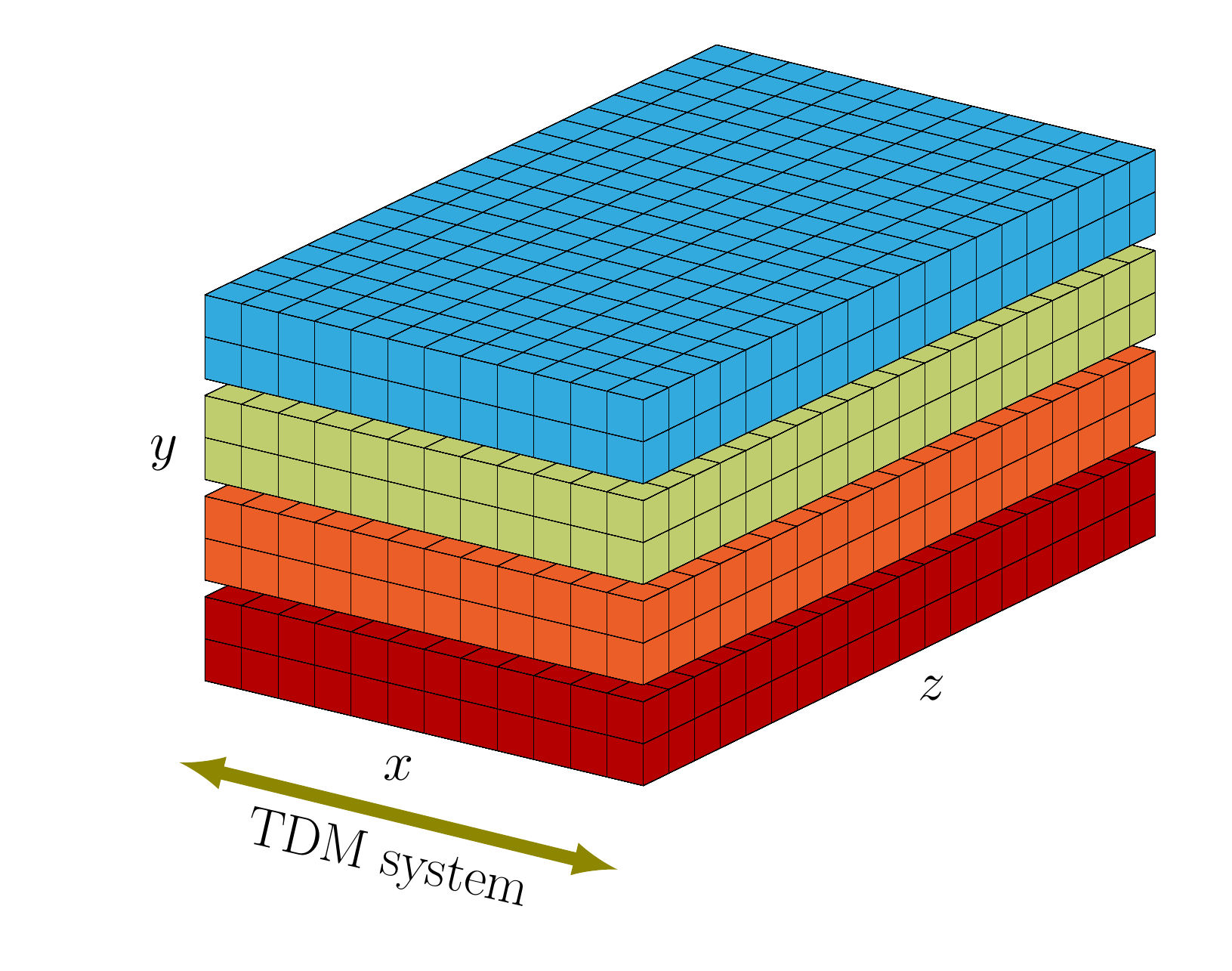}
        \subcaption{$x$ direction TDM sytems}
        \label{fig:x_tdma}
    \end{subfigure}
    \hspace{0.00\textwidth}
    \begin{subfigure}{0.32\textwidth}
        \centering
        \includegraphics[page=3, width=\linewidth]{figs/image2.pdf}
        \subcaption{$z$ direction TDM sytems}
        \label{fig:z_tdma}
    \end{subfigure}
    \hspace{0.00\textwidth}
    \begin{subfigure}{0.32\textwidth} 
        \centering
        \includegraphics[page=5, width=\linewidth]{figs/image2.pdf}
        \subcaption{$y$ direction TDM sytems}
        \label{fig:y_tdma}
    \end{subfigure}
    \caption{1D slab-type decomposition into four subdomains along the $y$ direction. The alignment direction of the tridiagonal systems is (a) $x$, (b) $y$, and (c) $z$.}
    \label{fig:1Dslab}
    
    \vspace{0.1cm} 

    \begin{subfigure}[b]{0.32\textwidth}
        \centering
        \includegraphics[page=2, width=\linewidth]{figs/image2.pdf}
        \subcaption{$x$ direction TDM sytems}
        \label{fig:x_tdma2}
    \end{subfigure}
    \hspace{0.00\textwidth}
    \begin{subfigure}{0.32\textwidth}
        \centering
        \includegraphics[page=4, width=\linewidth]{figs/image2.pdf}
        \subcaption{$z$ direction TDM sytems}
        \label{fig:z_tdma2}
    \end{subfigure}
    \hspace{0.00\textwidth}
    \begin{subfigure}[b]{0.32\textwidth}
        \centering
        \includegraphics[page=6, width=\linewidth]{figs/image2.pdf}
        \subcaption{$y$ direction TDM sytems}
        \label{fig:y_tdma2}
    \end{subfigure}
\caption{
Workload distribution of tridiagonal matrix systems when using four GPUs. When the TDM systems are aligned in the $z$- or $x$-direction, the matrix systems can be evenly distributed across the GPUs, as illustrated in (a) and (b). In contrast, when the data in TDM system are aligned in the $y$-direction, all four GPUs must cooperate to solve each TDM system, as shown in (c).
}

\vspace{0.1cm} 

\end{figure*}

However, TDM systems can be solved in multi-GPU systems, even when the data of each TDM system are distributed across multiple GPUs. 
In a one-dimensional (1D) slab-type decomposition along the $y$-direction, where the TDM systems are also aligned in the same direction (see Fig.\ref{fig:y_tdma}), the data within each TDM system are distributed across multiple GPUs (see Fig.~\ref{fig:y_tdma2}), making it impossible to solve a complete TDM system on a single GPU. A domain transpose is a commonly used solution that redistributes the data, ensuring that each TDM system becomes fully localized within a single GPU. This data rearrangement strategy is also widely adopted in FFT-based Poisson solvers and has been implemented in both GPU- and CPU-based flow solvers such as \citep{zhu2018afid, costa2021gpu, lee2010debunking, lee2013petascale}. As demonstrated by \citet{zhu2018afid}, this approach achieves good scalability up to thousands of GPUs when using a 2D pencil-type decomposition. However, these approaches require all-to-all communication for the domain transpose, and these solvers consistently report that this communication overhead remains a major bottleneck in the overall runtime.

Another approach to implementing TDMA on multi-GPU systems is through divide-and-conquer methods.
Instead of rearranging the TDM system via global transpose, divide-and-conquer methods partition a TDM system into smaller sub-systems and distribute the reduced workload across multiple GPUs.
In multi-GPU implementations of the divide-and-conquer method for parallel TDMA, each processor is mapped to an entire GPU, in contrast to CPU-based implementations, where processors are mapped to CPU cores, or single-GPU implementations, where they are mapped to threads or warps.
Due to this distinct characteristic, although most parallel TDMA algorithms fall under the divide-and-conquer category, only a subset of these methods \citep[e.g.,][]{sun1989parallel, mattor1995algorithm, polizzi2007spike} has been adopted for multi-GPU implementations, as seen in both standalone TDMA solvers~\citep{chang2012scalable, yang2023pascal_tdma} and integrated flow solvers~\citep{ha2021multi}.
Due to this distinct characteristic, although most parallel TDMA algorithms fall under the divide-and-conquer category, only a some of these methods \citep[e.g.,][]{sun1989parallel, mattor1995algorithm, polizzi2007spike} has been adopted for multi-GPU implementations, as seen in both standalone TDMA solvers~\citep{chang2012scalable, yang2023pascal_tdma} and integrated flow solvers~\citep{ha2021multi}.
This approach exhibits high performance by eliminating the need for global all-to-all communication.
However, the scalability of divide-and-conquer TDMA becomes critical when applied to FFT-based Poisson solvers, which require FFT operations in two spatial directions.
To avoid communication overhead from all-to-all transposes, only one-dimensional (1D) slab-type domain decomposition is permitted. Consequently, 2D and 3D decompositions—often used to improve scalability—are not applicable, making a highly scalable TDMA algorithm essential in this context.
Notably, several studies have sought to enhance the scalability of conventional divide-and-conquer TDMA methods.
In particular, recent work by \citet{kim2021pascal_tdma} improved CPU performance by evenly distributing intermediate TDM systems to reduce communication overhead, and this approach was later extended to GPU architectures by \citet{yang2023pascal_tdma}.

Despite these enhancements, little attention has been paid to sequential structure of divide-and-conquer method as used in multi-GPU systems, which limits GPU utilization.
Each TDM system is solved through multiple computation and communication stages in strict sequence, limiting the potential to exploit GPU features.
To address this limitation, This study proposes an algorithm that pipelines multiple communication/computation stages to hide most of the non-scalable portion of the divide-and-conquer TDMA.
To achieve this, three consecutive TDM systems are processed concurrently, allowing the communication and low-occupancy computation stages of one TDM system to overlap with the computation stages of the others. 
More specifically, the proposed method incorporates two forms of overlap.
First, communication–computation overlap hides communication latency behind computation, which proves highly effective when communication time is smaller than computation time—a condition typically satisfied in practice.
Second, concurrent kernel execution improves computational efficiency by executing small, low-occupancy TDM systems—arise in efficient divide-and-conquer TDMA—in parallel with larger ones.
This strategy reduces most of the non-scalable runtime, except during the terminal underfilled phase of the pipeline.
To mitigate the negative impact of the terminal phase and the performance loss arising from solving small main TDM systems, the batch size is optimized in advance through pre-computation.

The remainder of this paper is organized as follows.
Section~2 provides an in-depth analysis of the divide-and-conquer method, highlighting its performance and scalability limitations.
Section~3 introduces the proposed TDMA, followed by Section~4, which discusses batch size optimization to reduce the remaining non-scalable component.
Section~5 describes how TDM systems arise in incompressible Navier–Stokes solvers, while Section~ 6 outlines additional performance enhancements applied to the overall flow solver.
 Section7 demonstrates the solver’s versatility and validation across various benchmark cases. Finally, Section~8 provides a comprehensive performance evaluation of the proposed method.


\section{In-depth Investigation of Divide-and-Conquer Methods}

\begin{figure*}[b]
    \centering
        \centering
        \includegraphics[page=1, width=0.6\linewidth]{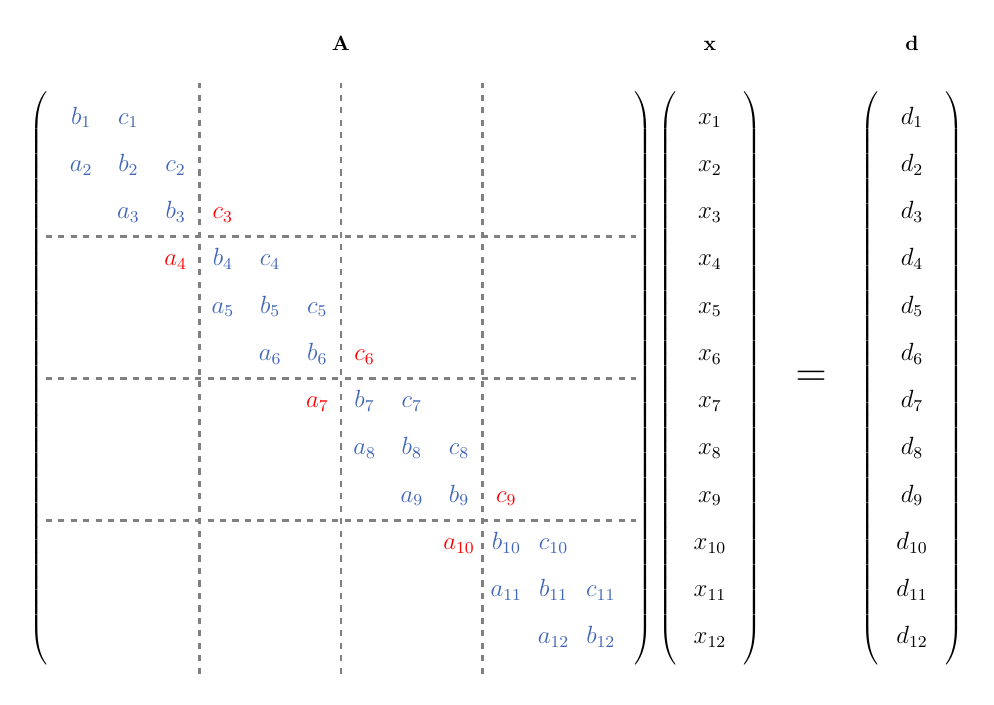}
        \caption{Example of a tridiagonal matrix system of size 12 used for illustration}
        \label{fig:subfig1}
\end{figure*}
\begin{figure*}[t]
    \centering
    \begin{subfigure}[b]{0.93\textwidth}
        \centering
        \includegraphics[page=2, width=\linewidth]{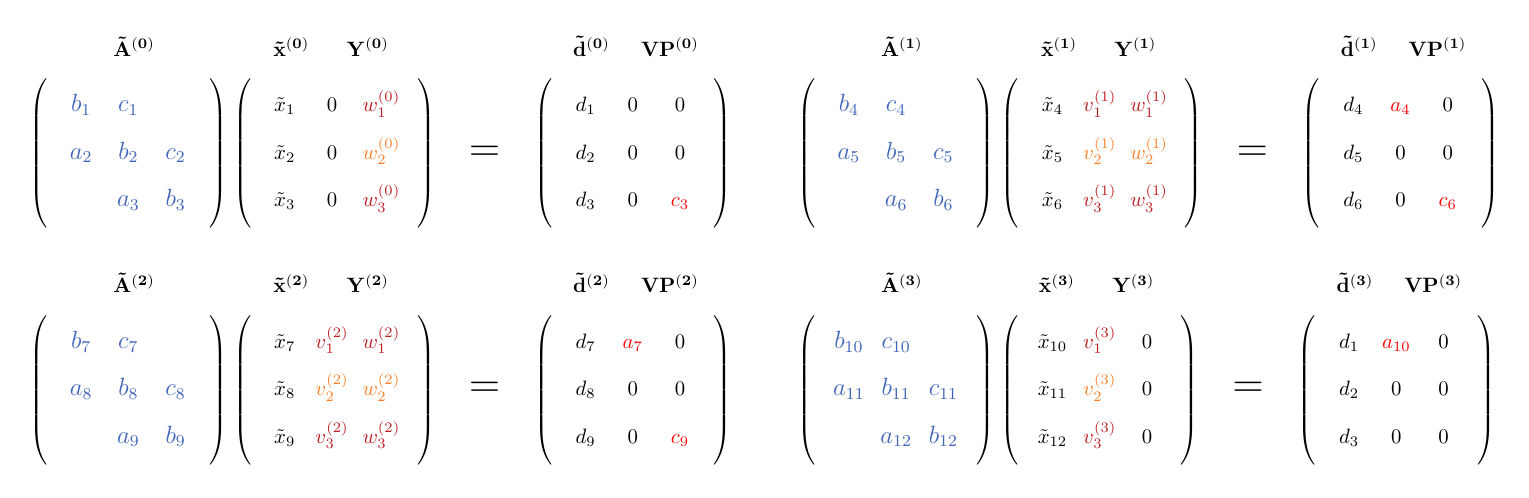}
        \caption{}
        \label{fig:subfig2}
    \end{subfigure}
    \vfill
    \begin{subfigure}[b]{0.55\textwidth}
        \centering
        \includegraphics[page=3, width=\linewidth]{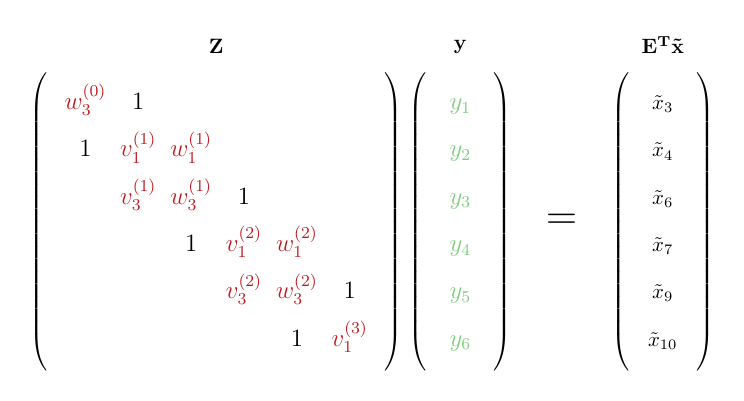}
        \caption{}
        \label{fig:subfig3}
    \end{subfigure}
    \vfill
    \begin{subfigure}[b]{0.93\textwidth}
        \centering
        \includegraphics[page=4, width=\linewidth]{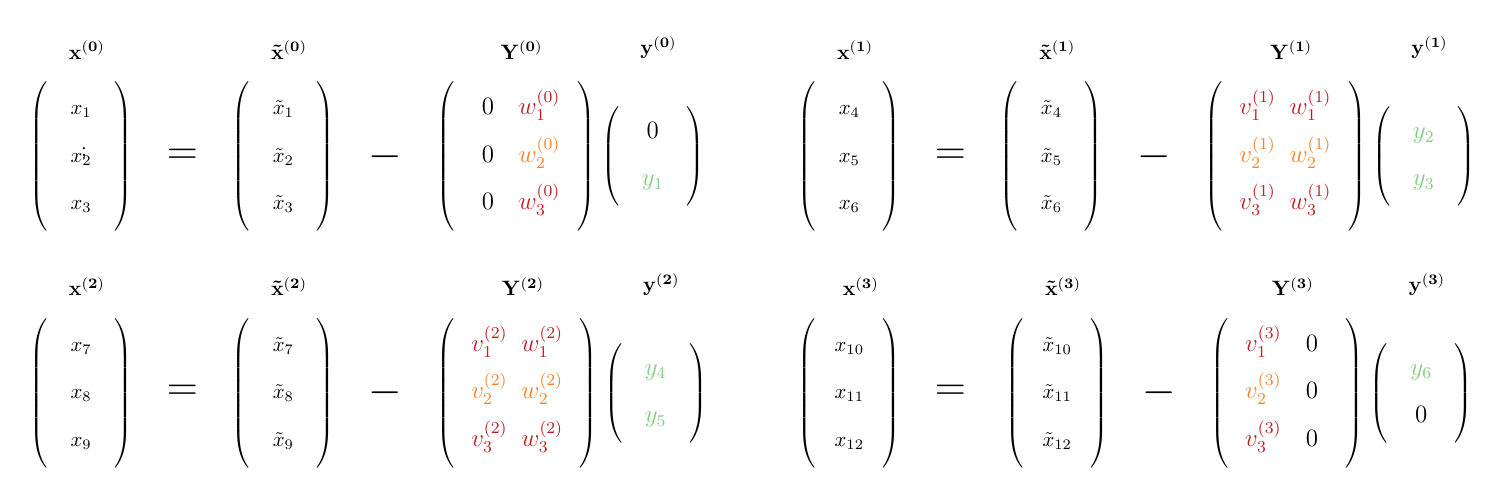}
        \caption{}
        \label{fig:subfig4}
    \end{subfigure}
    \caption{
        Illustration of the PPT method applied to a tridiagonal matrix system of size 12 using four processors.  
        (a) Each processor solves its assigned decoupled submatrix independently.  
        (b) A reduced interface system is assembled and solved by a designated processor.  
        (c) The complete solution is reconstructed using results from steps (a) and (b).
    }
    \label{fig:subfig_tot}
\end{figure*}

\subsection{General Structure of Divide-and-Conquer Methods}

Although a variety of divide-and-conquer strategies have been proposed for solving TDM systems, most follow a common structural pattern.
This section analyzes that structure using the PPT algorithm as a representative example. Originally introduced by~\citet{sun1989parallel} and later adapted for GPUs by~\citet{ha2021multi}, the PPT algorithm forms the foundation of the current implementation. A brief summary is provided here based on a TDM system of size 12 (Fig.~\ref{fig:subfig1}); detailed explanations can be found in the cited works.
Figure~\ref{fig:subfig_tot} illustrates the parallel solution procedure using four processors. The algorithm comprises five stages: Stages1, 3, and 5 involve computation, while Stages2 and 4 correspond to inter-processor communication. 

\begin{enumerate}
\item \textbf{Solving the decoupled TDM system} (Fig.\ref{fig:subfig2}): Each processor independently solves its assigned portion of the decoupled TDM system.
\item \textbf{Data gathering}: Each processor extracts interface-related data, which a designated processor then gathers from all others.
\item \textbf{Solving the reduced TDM system} (Fig.\ref{fig:subfig3}): The designated processor assembles and solves the reduced interface system (This system was previously referred to as the intermediate TDM system for clarity, but is hereafter termed the reduced TDM system.).
\item \textbf{Data scattering}: The solution to the reduced TDM system is broadcast to all processors.
\item \textbf{Final correction} (Fig.~\ref{fig:subfig4}): Each processor updates its local solution using the received data.
\end{enumerate}

\begin{figure} [htbp]
   \begin{center}
    \includegraphics[width=0.55\linewidth]{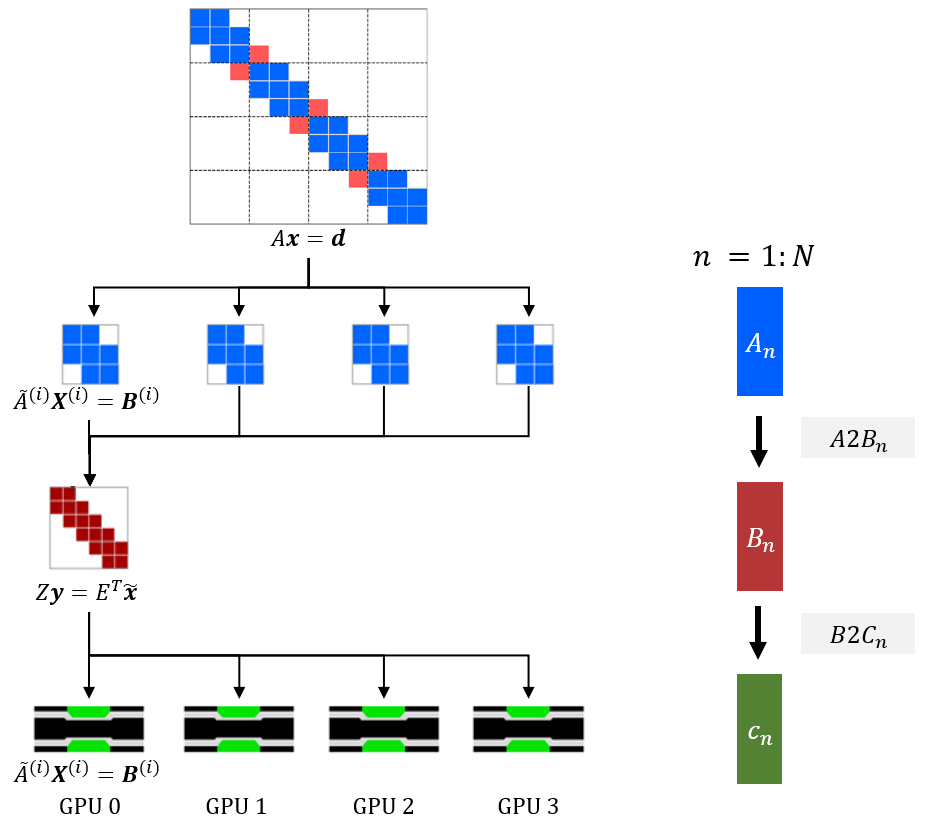} 
    \caption{Schematic of the PPT method, a divide-and-conquer approach. The algorithm consists of three main computation stages—Stage~A, Stage~B, and Stage~C—and two inter-processor communication stages—Stage~A2B and Stage~B2C—as labeled in the right-hand diagram and referenced throughout this paper.}
    \label{fig:PPT_schematic}
   \end{center}
\end{figure}

Figure~\ref{fig:PPT_schematic} summarizes the five stages, with Stage~A, B, and C representing the computation phases and Stage~A2B and B2C denoting the communication phases. For each of the $N$ TDM systems in a 3D domain, the loop $n = 1{:}N$ executes these steps. Algorithm~\ref{algorithm0} outlines the detailed logic.

\subsection{Time Complexity of Each Stage}

Divide-and-conquer methods define the global communication and computation structure across multiple GPUs, but rely on separate solvers to handle the reduced or decoupled TDM systems within each GPU.  
This study employs the CUDA TDMA library cuSPARSE to efficiently compute these TDM systems.
For sufficiently large TDM systems, cuSPARSE maintains linear computational complexity, ensuring scalable performance.

Let $L$ denote the size of the original TDM system before decoupling, and $p$ the number of processors. Stage A solves a TDM system of size $L/p$, while Stage C performs a correction step requiring approximately $4L/p$ floating-point operations (FLOPs).  
Both stages exhibit scalable performance, with execution time proportional to $L/p$.
In contrast, Stage B solves a reduced TDM system of size $2p - 2$, and the two communication stages—A2B and B2C—involve data transfers of $6p$ and $2p$ elements, respectively.  
These components exhibit increasing cost with processor count and are therefore classified as non-scalable stages.  

When $L/p \gg 2p - 2$, the scalable computation stages dominate and the contribution from non-scalable stages is negligible. 
However, when $2p - 2$ approaches $L/p$, the computational and communication costs from Stage~B and the two communication stages begin to dominate total execution time. 
Hence, divide-and-conquer methods require $L \gg p^2$ to maintain scalability, which limits their applicability in large-scale simulations.

\subsection{Scalability Limitations of Divide-and-Conquer Methods on Multi-GPU Architectures}
Two key factors constrain the scalability of divide-and-conquer methods like PPT on GPUs.
First, Stage~B introduces a bottleneck because only a single GPU performs the computation while all others remain idle. 
To alleviate this imbalance, the approach of evenly distributing the intermediate TDM systems across GPUs, as proposed by \citet{kim2021pascal_tdma}, effectively resolves this issue when solving multiple systems.

\begin{figure}[t]
   \begin{center}
    \vspace{0.01\textwidth}
    \begin{subfigure}[b]{1.0\textwidth}
        \centering
        \includegraphics[page=1, width=\linewidth]{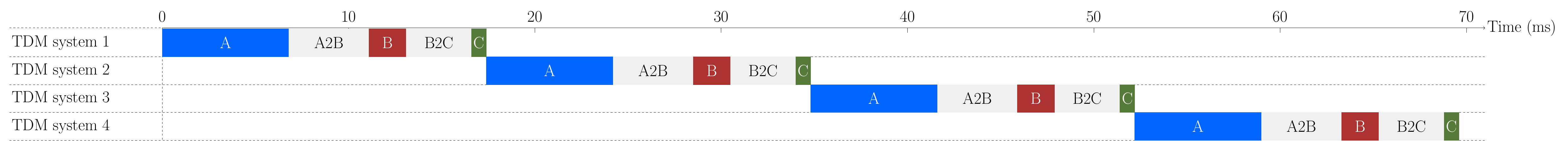}
        \caption{\textbf{PPT-v0}: Original sequential Algorithm.}
        \label{fig:original method}
    \end{subfigure}
    \\

    \vspace{0.01\textwidth}
    \begin{subfigure}[b]{1.0\textwidth}
        \centering
        \includegraphics[page=2, width=\linewidth]{figs/pipeline.pdf}
        \caption{\textbf{Intermediate pipelining}: Computation-communication overlap}
        \label{fig:proposed method}
    \end{subfigure}
    \\
    
    \vspace{0.01\textwidth}
    \begin{subfigure}[b]{1.0\textwidth}
        \centering
        \includegraphics[page=3, width=\linewidth]{figs/pipeline.pdf}
        \caption{\textbf{Final pipelining}: Computation-communication overlap \& concurrent kernel execution}
        \label{fig:proposed method2}
    \end{subfigure}
    \vfill
    
    \end{center}
    \caption{
    Pipelined execution of Stages A, A2B, B, B2C, and C across multiple TDM systems is illustrated, comparing the original sequential method (a) with the proposed pipelined strategies (b, c). The stage-wise execution times for \textbf{pipelined-v2} with a batch size of 32 on 64 GPUs, as reported in Table~2, are used as a representative example. In (b),  computation-communication overlap hides communication latency. In (c), concurrent kernel execution hides the performance loss associated with solving small TDM systems in Stage~B. When Stages~A and B are overlapped, their execution times are set to increase slightly, from 0.68 ms to 0.73 ms, and from 0.20 ms to 0.25 ms.
    }
\end{figure}

Second, the original sequential execution of stages (A, A2B, B, B2C, C) leads to underutilization of both communication and computation resources.
Resolving this inefficiency is a core objective of the proposed method.
Although data dependencies prevent pipelining within a single TDM system, solving multiple independent systems enables overlap between communication and computation stages across systems.
Additionally, concurrent execution of computation stages improves throughput by allowing multiple TDM systems to be processed in parallel.
In particular, performance degradation caused by low GPU occupancy during Stage~B—where small reduced TDM systems are solved—can be mitigated by overlapping this stage with other computation tasks via concurrent kernel execution.
The effectiveness of these techniques is confirmed through simple performance tests and aligns with the guidance presented in the “Parallelism with Streams” section of the cuSPARSE TDMA library\citep{nvidia2024cusparse}.
These strategies form the basis for the pipelined optimization described in the next section (see Fig.\ref{fig:proposed method} and Fig.\ref{fig:proposed method2}), where a more detailed explanation is provided.

\label{sec:sample1}

\section{TDMA Optimization Strategy}

Building on the PPT method~\cite{sun1989parallel}, this section presents an enhanced TDMA solver designed to efficiently handle multiple TDM systems in multi-GPU environments. For clarity, the original PPT algorithm is referred to as \textbf{PPT-v0} throughout this paper. The following subsections describe the optimization strategies that address two scalability bottlenecks identified in PPT-v0, as discussed in the previous section.

\subsection{Even distribution of the Stage~B workload across all processors}

The first scalability bottleneck arises from the idling of $p - 1$ processors during Stage~B. By adopting the modified divide-and-conquer strategy proposed by \citet{kim2021pascal_tdma}, this workload is evenly distributed across all GPUs, alleviating the bottleneck in both computation and communication related to Stage~B.

To achieve this distribution, the communication routines \texttt{MPI\_Gather} (Stage~A2B) and \texttt{MPI\_Scatter} (Stage~B2C) are replaced with \texttt{MPI\_Alltoall}, enabling all processors to participate equally in the data exchange.
Although the total communication volume remains unchanged, \texttt{MPI\_Alltoall} balances the data transfer workload across GPUs and significantly reduces overall communication time.
On the computation side, each processor solves a portion of the reduced batched TDM system, keeping the matrix size per GPU nearly constant as the number of GPUs increases. Further implementation details are available in~\citet{kim2021pascal_tdma}.

The baseline TDMA solver, before applying pipelining, is based on the GPU-parallel PPT algorithm developed by~\citet{ha2021multi}, augmented with the even-distribution strategy proposed in~\citet{kim2021pascal_tdma}.
As shown in the \texttt{pipelined-v2} results in Table~\ref{tab:results1}, the baseline implementation still suffers from high non-scalable runtime, with the combined cost of communication and Stage~B computations accounting for 129\% and 74\% of the scalable stage time on 64 GPUs for different batch sizes, respectively.
The pipelined method introduced in the next subsection effectively hides most of this remaining overhead.

\subsection{Pipelining via Communication–Computation Overlap and Concurrent Kernel Execution}

The second scalability bottleneck in PPT-v0 arises from the strictly sequential execution of all stages. 
While GPUs support concurrent execution through pipelining, this capability remains unused under such a sequential structure. 
However, solving multiple independent TDM systems enables overlapping of stages across different systems. This overlapping allows concurrent execution of computation and communication stages, improving GPU utilization and overall performance.

Computation-communication overlap is a common strategy in GPU-based solvers to hide communication latency. For instance,\citet{nguyen2019gpu} and\citet{ravikumar2021extreme} apply this technique during ghost cell exchanges for Jacobi iteration methods and for solving Poisson equations using spectral methods, respectively. Leveraging a similar strategy, this implementation introduces an intermediate pipelining method to hide communication stages, as shown in Fig.~\ref{fig:proposed method}.

Beyond communication, the pipelining strategy also hides performance degradation caused by low GPU occupancy through concurrent kernel execution (CKE). 
Stage~B, which solves significantly smaller TDM systems than Stage~A, frequently suffers from low occupancy. 
Typically, the TDM size ratio between Stage~A and Stage~B is approximately $L/p : 2$. 
For more information on this ratio, refer to the performance analysis section, where the exact TDM sizes used in the test cases are provided.
Therefore, although TDM systems are batched to mitigate GPU low occupancy, avoiding low performance in Stage B is practically impossible. To fully resolve this, the batch size would need to be excessively large, reducing the number of batched TDM systems and ultimately degrade pipelining efficiency. 

Instead, the final pipelining strategy (Fig.\ref{fig:proposed method2}) resolves this issue using two non-default CUDA streams: one for Stage~B and another for Stages A and C. This allows Stage~B to execute concurrently with other computation stages (or to overlap with them), mitigating low occupancy.
Empirical results confirm the effectiveness of this strategy.
On Tesla P100 (8 GPUs) with 1 billion grid cells, the total execution time for computation stages A, B, and C was observed to be shorter than the sum of the individual execution time of each stage, without any communication stages. Additionally, our results tested on A100 (64 GPUs) further confirm this. 
In some tested cases, the total execution time is even smaller than the sum of the individual execution times for all computation stages, despite the fact that total execution includes overlapped communication stages.

As a result, the overlap hides both performance loss of low-occupancy computation and communication latency within scalable stages. 
The method solves three consecutive TDM systems in parallel, aligning non-scalable stages with computation-heavy scalable stages to maximize pipeline utilization.
It should be noted that during the pipeline's end phases, underutilization might occur. As shown in end phase of Fig.\ref{fig:proposed method}, Since execution time of Stage~B and Stage~C are generally shorter than Stages A2b and B2C, there's some empty region for computation that cannot fully hide communication latency. This overhead becomes negligible as the number of systems (iterations) increases. To achieve full computation-communication overlap, the duration of the communication stage must be significantly short. This condition is generally satisfied when using a small number of GPUs. However, as the GPU count increases and communication time exceeds the duration of Stage~B and Stage~C, some communication latency remains exposed. Nonetheless, when the number of iterations $N$ is sufficiently large, the impact of this residual overhead becomes negligible over many iterations.

\begin{figure}[htbp]
\centering
\begin{minipage}[t]{0.48\textwidth}
\begin{algorithm}[H]
\fontsize{10}{12}\selectfont
\caption{Original Sequential Method}
\begin{algorithmic}[1]
\State \textbf{Problem} : $\boldsymbol{A}\boldsymbol{x} = \boldsymbol{d}$  ($\textit{N}$ times)
\State \textbf{Input} : \ \ \ \ \ \ $\boldsymbol{a}, \boldsymbol{b}, \boldsymbol{c}, \boldsymbol{d}$
\State \textbf{Output} : \ \ \ $\boldsymbol{x}$
\State
\State do \ $n = 1\!:\!N$
\State \ \ \ \textit{\textbf{Stage A}}
\State \ \ \ \ \ \ Solve decoupled matrix system $\mathbf{\tilde{Ax}=\tilde{d}}$
\State \ \ \ \ \ \ Save data in send\_A 
\State \ \ \ \textit{\textbf{Stage A2B}}
\State \ \ \ \ \ call MPI\_Gather (send\_A, recv\_B)
\State \
\State \ \ \ \textit{\textbf{Stage B}}
\State \ \ \ \ \ \ Get data from recv\_B 
\State \ \ \ \ \ \ Solve reduced matrix system $\mathbf{Zy}=\mathbf{E}^T\mathbf{\tilde{x}}$
\State \ \ \ \ \ \ Save data in send\_B 
\State \ \ \ \textit{\textbf{Stage B2C}}
\State \ \ \ \ \ \ call MPI\_Scatter (send\_B, recv\_C)
\State \
\State \ \ \ \textit{\textbf{Stage C}}
\State \ \ \ \ \ \ Get data from recv\_C 
\State \ \ \ \ \ \ Get final results 
$\mathbf{x}=\mathbf{\tilde{x}} - \mathbf{Yy}$
\State enddo
\end{algorithmic}
\label{algorithm0}
\end{algorithm}
\end{minipage}
\hfill
\begin{minipage}[t]{0.48\textwidth}
\begin{algorithm}[H]
\fontsize{10}{12}\selectfont
\caption{Proposed Pipelined Method}
\begin{algorithmic}[1]
\State \textbf{Problem} : $\boldsymbol{A}\boldsymbol{x} = \boldsymbol{d}$  ($\textit{N}$ times)
\State \textbf{Input} : \ \ \ \ \ \ $\boldsymbol{a}, \boldsymbol{b}, \boldsymbol{c}, \boldsymbol{d}$
\State \textbf{Output} : \ \ \ $\boldsymbol{x}$
\State
\State do $\textcolor{red}{n} = 1\!:\!N\!+\!2$
\State \ \ \textit{\textbf{Stage A}}
\State \ \ \ \ Solve decoupled matrix system ($\textcolor{red}{n}$, \textcolor{blue}{stream1}) 
\State \ \ \ \ Save data in send\_A ($\textcolor{red}{n}$, \textcolor{blue}{stream1}) 
\State \ \ \textit{\textbf{Stage A2B}}
\State \ \ \ \ call MPI\_Alltoall (send\_A, recv\_B, $\textcolor{red}{n-1}$) 
\State \ \ \ \ call (cudaStreamSynchronize(\textcolor{blue}{stream2})) 
\State \ \ \textit{\textbf{Stage B}}
\State \ \ \ \ Get data from recv\_B ($\textcolor{red}{n-1}$, \textcolor{blue}{stream2}) 
\State \ \ \ \ Solve reduced matrix system ($\textcolor{red}{n-1}$, \textcolor{blue}{stream2})
\State \ \ \ \ Save data in send\_B ($\textcolor{red}{n-1}$, \textcolor{blue}{stream2}) 
\State \ \ \textit{\textbf{Stage B2C}}
\State \ \ \ \ call MPI\_Alltoall (send\_B, recv\_C, $\textcolor{red}{n-2}$)
\State \ \ \ \ call cudaStreamSynchronize(\textcolor{blue}{stream1}) 
\State \ \ \textit{\textbf{Stage C}}
\State \ \ \ \ Get data from recv\_C ($\textcolor{red}{n-2}$, \textcolor{blue}{stream1}) 
\State \ \ \ \ Get final results ($\textcolor{red}{n-2}$, \textcolor{blue}{stream1}) 
\State enddo
\end{algorithmic}
\label{algorithm1}
\end{algorithm}
\end{minipage}
\end{figure}

\subsubsection{Algorithm of the proposed method}

In Algorithm~\ref{algorithm1}, $a$, $b$, and $c$ represent the sub-diagonal, main diagonal, and super-diagonal entries of the tridiagonal matrix $A$, respectively. 
The algorithm solves $N$ independent TDM systems by concurrently executing of three consecutive TDM systems at once, indexed as $n-2$, $n-1$, and $n$. 
If any index falls outside the valid range $[1, N]$, the corresponding operation is skipped.
These boundary checks are omitted from the pseudocode for clarity. 
Figure~\ref{fig:proposed method2} provides a visual illustration of how the algorithm operates.

Compared to Algorithm~\ref{algorithm0}, the proposed algorithm uses three CUDA streams—one default and two non-default—for asynchronous execution of: (1) scalable computation stages, (2) non-scalable computation stage, and (3) communication stages.
The default stream handles communication via CUDA-aware MPI \texttt{MPI\_Alltoall}. Non-default stream1 executes Stages A and C, while non-default stream2 handles Stage~B. Synchronizations using \texttt{cudaStreamSynchronize} ensure sequential consistency for stages within the same TDM system.

This design allows all communication stages to overlap with computation and lets the non-scalable Stage~B execute in parallel with scalable stages A and C. The implementation was verified for correctness and integrated into the flow solver of \citet{ha2021multi}. The solution accuracy remained consistent, and performance evaluation confirmed that most of the communication latency and impact of low occupancy are successfully hidden.

\section{Batch Size Optimization}

Optimizing batch size is essential to fully realize the performance potential of the pipelined TDMA. 
This section presents a detailed analysis aimed at avoiding the critical bottleneck of pipelined TDMA through batch-level tuning.

\subsection{Prior Research on Batched TDM Systems}

Small TDM systems perform poorly on GPUs due to low occupancy. For example, Fig.~6 of \citet{ha2021multi} illustrates the performance degradation of the CUDA built-in TDMA solver \texttt{cusparseDgtsv2} when solving small TDM systems.

To address this limitation, prior studies have improved GPU performance by aggregating multiple small TDM systems into larger batched systems~\citep{yang2023pascal_tdma, valero2018cuthomasbatch, chang2012scalable}. However, most of these studies do not offer explicit strategies for selecting batch sizes. \citet{ha2021multi} proposed an empirical approach that tests various batch sizes before the main computation. Their estimation, however, considers only Stage A and ignores communication (Stages A2B and B2C), which is critical for time estimation. This omission causes significant deviations from actual execution time, particularly at large GPU counts where communication costs become substantial.

\subsection{Enhanced Batch Optimization Method}

The proposed batch size optimization improves time estimation by incorporating both computation and communication costs. Total execution time is predicted across a range of batch sizes—from near 1 to a sufficiently large value—and the batch size that yields the shortest estimated execution time is chosen as optimal. Since evaluating a wide range of batch sizes could introduce substantial pre-computation time, the method is designed to balance accuracy with minimal overhead.

\subsubsection{Batched TDM System Construction}

Batched TDM systems are constructed to improve GPU occupancy by grouping multiple small systems into larger composite systems. This approach reduces kernel launch overhead and increases GPU occunpancy, enhancing overall performance on GPU architectures.

In a 3D domain where systems are aligned along the $y$-direction, the simulation requires solving $n_x \times n_z$ TDM systems of size $n_y$ (see Fig.~\ref{fig:y_tdma}). These are first grouped along the $x$-direction, forming $n_z$ TDM systems of size $n_x \times n_y$. To further enhance GPU occupancy, systems are aggregated along the $z$-direction using a user-defined parameter $batch_z$, yielding $(n_z / batch_z)$ batched TDM systems of size $n_x \times n_y \times batch_z$. If $n_z$ is not divisible by $batch_z$, an additional smaller system is created for the remainder.

\subsubsection{Three Factors Determining Batch Size}

Three primary factors govern the selection of the optimal batch size, and the first two are incorporated into the time estimation procedure.
First, a larger batch size leads to larger TDM systems and improves GPU occupancy.
Stage~C involves minimal computation due to simple matrix operations, whereas StagesA and B solve TDM systems and dominate the total runtime and require sufficiently large batches to ensure high GPU utilization.
Among these, Stage~A contributes the most to the overall execution time because of its larger matrix size. Therefore, maximizing GPU occupancy in Stage~A is critical.

Second, small batch sizes reduce underfilled pipeline effects. As shown in Fig.\ref{fig:proposed method}, the end of the pipeline underutilizes some computation stages, exposing communication latency. Larger batch sizes reduce the total number of batched systems, amplifying this effect. To mitigate this issue, the batch size must be small enough to maintain a sufficiently large number of batched TDM systems. This behavior contrasts with the results of \citet{ha2021multi}, where performance plateaus after a threshold batch size. The proposed pipelined method remains sensitive to batch count and requires careful tuning. When communication time is short relative to the execution time of Stages B and C, the impact of this pipeline inefficiency becomes negligible.

Third, memory constraints must be considered. The pipelined execution method increases temporary memory demands because it processes three batched systems concurrently. The total memory footprint includes: (1) 3D domain data, (2) temporary storage for the TDM solver, and (3) internal buffers for \texttt{cusparseDgtsv2}. Among these, temporary storage grows threefold relative to the original method. For example, solving a $2048 \times 256 \times 2048$ complex-number domain on an NVIDIA A100 GPU with batch size 32 requires approximately 17GB for domain data, 6GB for cuSPARSE buffers, and 6GB (up from 2GB) for temporary storage. This 4~GB increase remains small compared to the total memory usage.

In summary, the batch size optimization process should be designed to minimize the combined impact of two performance loss:  
(1) performance degradation when solving small TDM systems in Stage~A, and  
(2) pipeline inefficiencies caused by underfilled stages at the end of execution.  
Although this approach increases temporary memory usage, the overhead remains negligible compared to the dominant memory cost of storing 3D field data.

\subsubsection{Estimating Execution Time to Determine the Optimal Batch Size}

To avoid the high cost of measuring the full execution time over the entire 3D domain, a simplified and theoretical time estimation algorithm is used. The average execution times for each stage are first measured individually, and then the total execution time is estimated based on a pipelined scheduling model, without solving all TDM systems.

This estimation does not aim to produce exact execution time, as high accuracy is not achievable using theoretical prediction based on pipelined scheduling model. In fact, in the performance tests, even when communication latency is shorter than the overlappable computation time, 100\% communication latency hiding is not always realized.

Instead, the focus lies on identifying the most effective batch size by comparing relative execution time trends. The estimation specifically targets the two dominant factors affecting performance: (1) low performance in Stage~A for small batch sizes, and (2) underutilization at the end of the pipeline due to incomplete overlap. For this purpose, the simplified pipelining structure shown in Fig.\ref{fig:proposed method} is used, instead of the more complex pipelining structure shown in Fig.\ref{fig:proposed method2}.

To evaluate whether neglecting the impact of low GPU occupancy in Stage~B affects the selection of the optimal batch size when using the simplified pipelining structure, tests were conducted across various GPU counts.
The theoretical case assumes that the performance loss from low occupancy in Stage~B is fully hidden by overlapping with other computation stages.
Under this assumption, Stage~B is expected to exhibit the same computation speed as Stage~A when accounting for the ratio of TDM matrix sizes.
Note that the actual execution time is not the focus of this evaluation, as the goal is to identify the optimal batch size rather than predict runtime accurately.
Therefore, if the relative performance trends across batch sizes remain consistent, the use of a simplified pipelining structure for time estimation is considered justified to accelerate pre-computation.
The largest relative execution time difference between the theoretically optimal batch size and the original optimal batch size using the simplified pipeline was found to be only 1.6\%, as observed at 16 GPUs in the strong scaling tests presented in the performance analysis section. Other configurations exhibitfed negligible error: 4 GPUs: 0.06\%, 8 GPUs: 0.05\%, 32 GPUs: 0.23\%, and 64 GPUs: 0.54\%. Based on these results, this estimation strategy is deemed sufficiently accurate for batch size selection.

\section{Occurrence of TDM Systems in Navier–Stokes Simulations}

This section describes how TDM systems naturally arise in the numerical solution of the incompressible Navier–Stokes equations and highlights their computational significance. The incompressible Navier--Stokes equations used in this study are:
\begin{align*}
\partial_i u_i &= 0, \\
\frac{\partial u_i}{\partial t} + u_j \partial_j u_i &= -\partial_i P + \frac{1}{Re} \partial_j^2 u_i,
\end{align*}
where $u_i$ denotes the velocity components, $P$ the pressure, and $Re$ the Reynolds number.
A semi-implicit fractional step method~\citep{chorin1967numerical} solves these equations through two sub-steps: a momentum update and a pressure projection. The first step computes an intermediate velocity field $\hat{\mathbf{u}}$ without imposing incompressibility. The second step enforces the divergence-free condition by solving a Poisson equation for pressure. The momentum equation applies the Crank--Nicolson method to treat the viscous term implicitly and a third-order low-storage Runge--Kutta scheme for explicit integration of the convective term. Further numerical details are available in \citet{ha2021multi, ha2018scalable}.

Simulations take place in a Cartesian coordinate system, with $x$, $y$, and $z$ denoting the streamwise, wall-normal, and spanwise directions, respectively. The domain is decomposed along the $y$-direction, as illustrated in Fig.~\ref{fig:y_tdma}. Grid spacing remains uniform in both $x$ and $z$ directions.

\subsection{Solving the Momentum Equation}
The spatial discretization uses second-order central finite-difference schemes on a staggered grid. 
The Alternating Direction Implicit (ADI) method reformulates the momentum equation as~\cite{ha2021multi}:
\[
(1 - A_x)(1 - A_y)(1 - A_z)(\hat{\mathbf{u}} - \mathbf{u}) = \mathbf{R}.
\]

Here, $(1 - A_x)$, $(1 - A_y)$, and $(1 - A_z)$ represent tridiagonal operators with small off-diagonal entries, resulting in diagonally dominant operators. 
The intermediate velocity $\hat{\mathbf{u}}$ is obtained by sequentially solving three such TDM systems.

Because the domain is decomposed along the $y$-direction, solving the $(1 - A_y)$ system requires a parallel TDMA. 
This implementation adopts the Parallel Diagonally Dominant (PDD) method~\citep{sun1989parallel}, which is designed for diagonally dominant TDM systems. 
The PDD algorithm minimizes communication overhead and maintains excellent scalability on GPUs, making it effective for large-scale momentum solves.

\subsection{Solving the Poisson Equation Using the Proposed Pipelined Algorithm}

To solve the Poisson equation, the algorithm applies Fourier transforms in the $x$ and $z$ directions, reducing the original equation to~\cite{ha2021multi}:
\[
\left( \frac{\delta^2}{\delta y^2} - k_l - k_m \right) \Phi = Q,
\]
where $k_l$ and $k_m$ denote the Fourier wavenumbers in the $x$ and $z$ directions. The equation becomes a TDM system in the $y$-direction.

Domain decomposition along the $y$-axis is chosen to support efficient FFT-based Poisson solves under the cuFFT library’s single-GPU restriction. This configuration allows each GPU to independently perform FFTs in $x$ and $z$ directions without inter-GPU communication. In contrast, decomposing along $x$ or $z$ would require global all-to-all transpositions, which impose significant communication overhead.

To solve the resulting TDM systems, the original solver used the Parallel Partition Tridiagonal (PPT) method~\cite{sun1989parallel}, which supports general TDM systems. However, PPT exhibits limited scalability on large multi-GPU systems. To address this limitation, the proposed pipelined algorithm enhances the PPT method with multi-GPU optimizations. This improved implementation achieves substantial performance gains on up to 64 GPUs and directly enhances the scalability and runtime efficiency of the entire flow solver, making it suitable for large-scale simulations.

\section{Flow Solver optimization}
This section presents additional optimizations beyond the TDMA improvements in the Poisson solver, aligning with the broader objective of developing a high-performance flow solver. As part of this effort, the Poisson solver is enhanced through an optimized variant, referred to as \textbf{Poisson-v1}, which reduces overall execution time without requiring additional memory by reusing buffers allocated for the momentum equation.
The momentum equation is also accelerated by modifying the TDM system configuration—increasing the size of the coefficient matrix $A$ in $Ax = b$—which leads to a meaningful performance gain.

Since the implementation builds upon the codebase of \citet{ha2021multi}, the comparisons made here emphasize the direct differences relative to their implementation.

\subsection{\textbf{Poisson-v1}: Resolving Non-Coalesced Memory Access in the Poisson Solver}

Although \citet{ha2021multi} addressed non-coalesced memory access in many components of their solver, inefficiencies remain, particularly during matrix transformations in the Poisson step.

In boundary layer simulations, their method first applies a Fast Fourier Transform (FFT) in the spanwise ($z$) direction, followed by a half-cosine transform in the streamwise ($x$) direction. After these transforms, the solver performs TDM computation in the wall-normal ($y$) direction, followed by inverse Fourier transforms in the $x$ and $z$ directions.

To minimize memory consumption, the FFT and half-cosine transforms are computed on 2D streamwise-spanwise planes, repeated along the $y$-direction. However, this introduces a memory layout mismatch: the transformed 2D data use $(k, i)$ index order, while the subsequent 3D array uses $(j, i, k)$ ordering in FORTRAN. As a result, directly mapping between these layouts leads to non-coalesced memory access, degrading performance.

This issue is mitigated by introducing a temporary 3D array with indexing $(k, i, j)$. Although non-coalesced access persists, 3D-to-3D transfers yield lower latency than inserting 2D data repeatedly into 3D arrays in a non-contiguous manner.
This solution requires no additional memory allocation since the temporary array shares memory with the existing momentum equation workspace. 
When applied to a 130-million-cell domain on an NVIDIA A100 GPU, this optimization achieved a 29\% speedup in combined forward and inverse FFT stages.

\subsection{Enhancement of the Momentum Equation Solver}
The ADI method consumes a substantial portion of the execution time, particularly at low GPU counts. it's due to 9 times of solving TDM system compared to the poisson solver, although only 3 of 9 require divide-and-conquer TDMA. therefore, optimizaint TDMA is important regardless of multi-gpu TDMA or single-gpu algorithm.

\subsubsection{Single-GPU TDMA Enhancements}
In the original implementation, batching the tridiagonal matrix (TDM) systems of the form $Ax = b$ transforms the operation from $(N, N) \cdot (N, 1) = (N, 1)$ to $(N, N) \cdot (N, M) = (N, M)$, where $M$ is the batch size. Since the coefficient matrix depends solely on grid spacing, all systems in the batch share the same matrix $A$, and thus the matrix size remains unchanged regardless of the batch size. {This formulation reduce memory requirements}; however, performance degrades when the coefficient matrix becomes too small—particularly when using \texttt{cusparseGtsv2} library.

To address this issue, the batched TDM system is expanded to $(NM_1, NM_1) \cdot (NM_1, M_2) = (NM_1, M_2)$, where $M = M_1 M_2$. This extension increases the size of the coefficient matrix. $M_1$ is chosen from $\{1, 2, 4, 8, 16\}$ when $M$ is divisible by those values. To avoid excessive pre-computation, the number of tested batch sizes $M$ is reduced.
{On 8 A100 GPUs at the Korea Supercomputing Center (KISTI), with a $1024 \times 128 \times 1024$ grid per GPU, this enhancement reduces TDMA execution time by 53.0\% along the $x$ direction and 46.5\% along the $y$ direction compared to the original implementation~\citep{ha2021multi}, resulting in an overall 29.2\% time reduction in the execution time of the momentum equation.} Since this optimization targets single-GPU execution, its effectiveness remains independent of GPU count.

\subsubsection{About {pipelining} methods for TMDA in momentum equation.}
Among the nine TDM systems in the ADI scheme, only three—those aligned with the wall-normal ($y$) direction—require multi-GPU TDMA at each time step. Since the $y$ direction does not impose periodic boundary conditions, its associated workload remains limited. As a result, only about one-fifth to one-fourth of the tasks for solving TDM systems in the momentum equations involve the $y$ direction.

The Parallel Diagonally Dominant (PDD) algorithm already provides excellent scalability by distributing the reduced matrix system in Stage~B and employing a simple, stable communication pattern. Due to its low communication overhead, applying a pipelined overlap strategy offers limited benefit and is therefore omitted from the PDD-based momentum solver. 
{
Even after hiding all communication latency within computation, tests on 8 A100 GPUs at the Korea Supercomputing Center (KISTI) with a $1024 \times 128 \times 1024$ grid showed only a 0.8\% performance gain. Because communication cost scales with the number of halo cells, further improvements may arise under strong scaling as GPU count increases; however, no additional benefit is expected under weak scaling.}

\section{Versatility and Validation of the Flow Solver}
The enhanced flow solver reproduces various canonical flow scenarios on a local system equipped with eight Tesla P100 GPUs, providing a total of 128~GB memory. As reported by \citet{ha2021multi}, this configuration supports simulations with up to approximately 1.4 billion grid cells. Despite the limited GPU count, the solver maintains robust performance across all test cases, demonstrating its versatility and efficiency. The high scalability of the proposed TDMA enables seamless extension to higher-resolution simulations in large-scale computing environments.

\subsection{Turbulent Channel Flow at $Re_\tau = 1000$}
\begin{figure*}[htbp]
    \centering
    \begin{subfigure}{0.65\textwidth}
        \includegraphics[width=\linewidth,trim=450 0 350 0, clip]{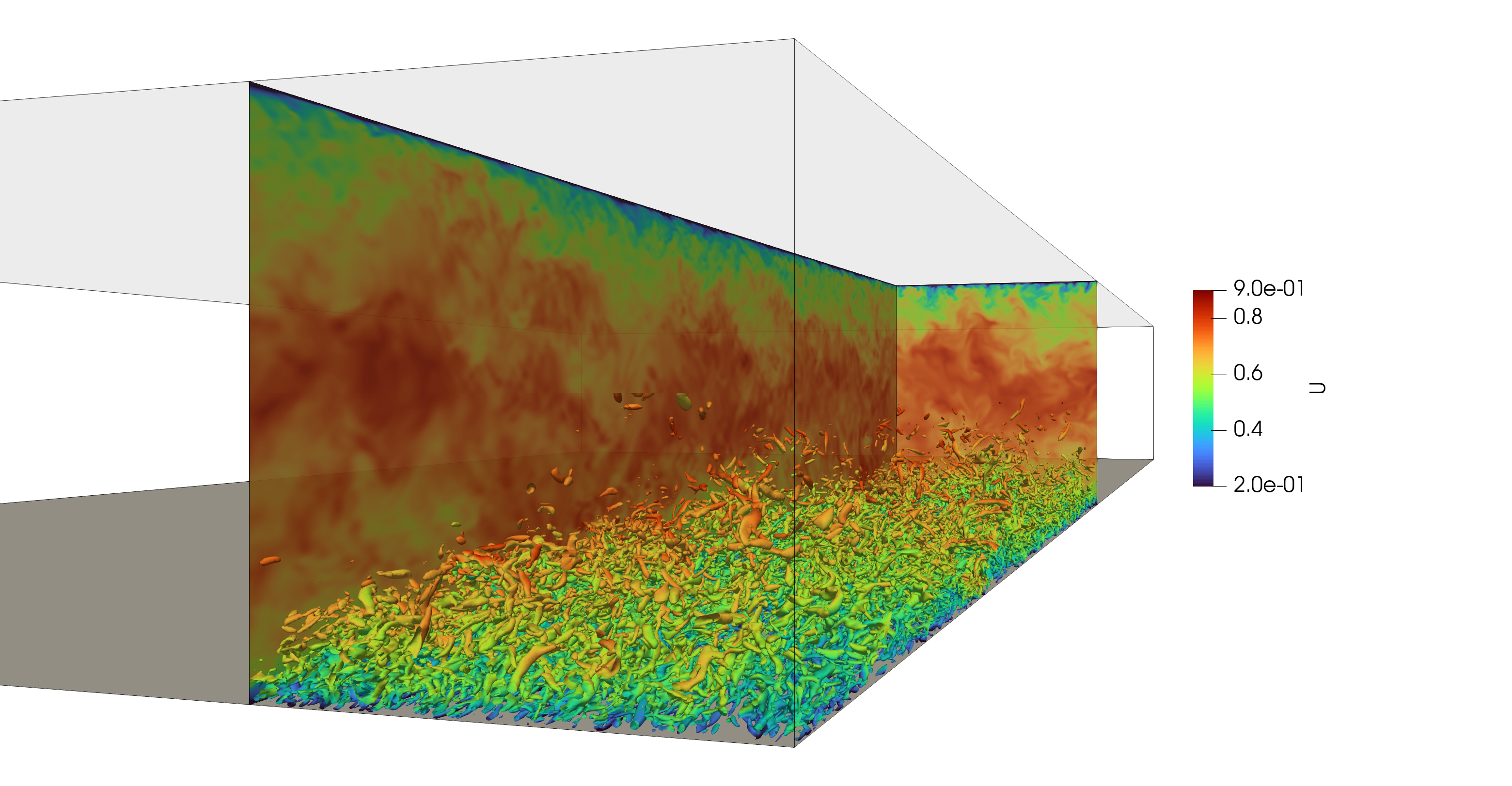}
        \label{fig:top}
        \vspace{-1.5em}
        \caption{}
    \end{subfigure}
    \par\vspace{0.5em}
    \centering
    \begin{subfigure}{0.38\textwidth}
        \centering
        \includegraphics[width=\linewidth,trim=0 0 0 0, clip]{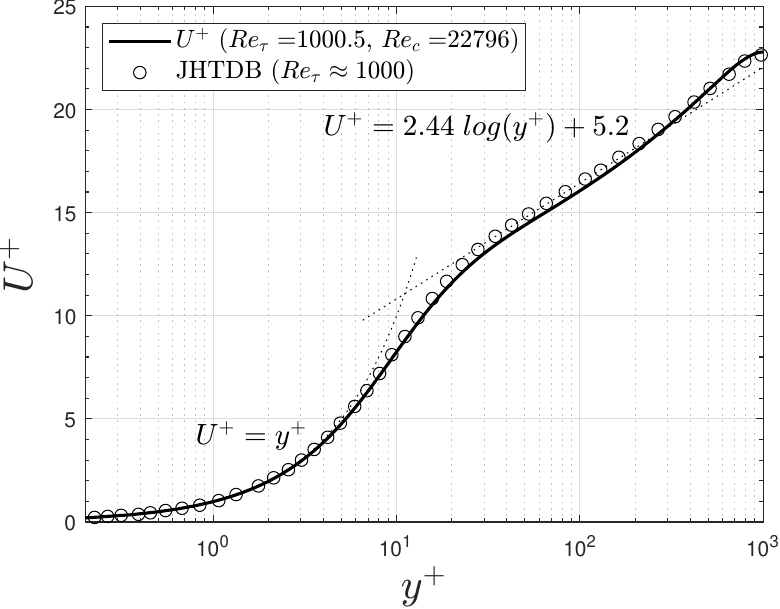}
        \label{fig:vortex1}
        \vspace{-1.5em}
        \caption{}
    \end{subfigure}
    \hspace{0.01\textwidth} 
    \begin{subfigure}{0.355\textwidth}
        \centering
        \includegraphics[width=\linewidth,trim=0 0 0 0, clip]{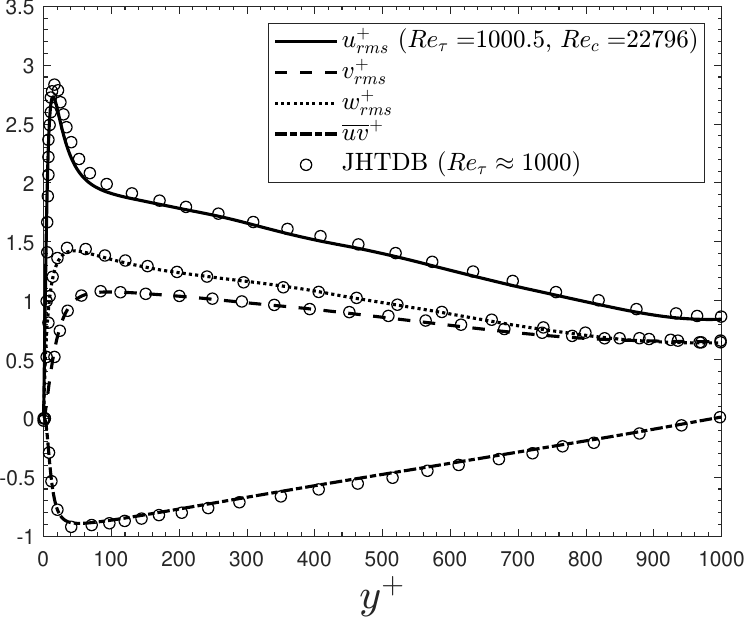}
        \label{fig:vortex2}
        \vspace{-1.5em}
        \caption{}
    \end{subfigure}
    \caption{{Turbulent channel flow at $Re_\tau \approx 1000$: (a) visualization of Q-criterion iso-surfaces in the lower half of a selected region, and comparisons of (b) mean velocity and (c) velocity fluctuation profiles.}}

    \label{fig:turbulentchannel}
\end{figure*}

The original boundary-layer flow solver is extended to simulate fully developed turbulent channel flow. Direct numerical simulation (DNS) at friction Reynolds number $Re_\tau = 1000$, corresponding to a bulk Reynolds number of $Re_b = {40000}$, is conducted on a grid of $(N_x, N_y, N_z) = (1536, 512, 1536)$ points over a domain of size $(L_x, L_y, L_z) = (6\pi h, 2h, 3\pi h)$, where $h$ denotes the channel half-height. The grid employs a hyperbolic tangent distribution in the wall-normal direction to enhance near-wall resolution. The grid spacing is {$\Delta x^+ = 12.3$, $\Delta z^+ = 6.14$}, and the first wall-normal spacing is {$\Delta y^+ = 0.138$}. Simulation results are compared against reference DNS data from the Johns Hopkins Turbulence Database (JHTDB~\citep{setjhu}). Statistical sampling spans {20{,}000} time steps, with each step requiring approximately 5.1 seconds. As shown in Fig.~\ref{fig:turbulentchannel}, the mean velocity profile adheres to the law of the wall, and the turbulence intensity profiles exhibit {good} agreement with the reference data.

\subsection{Bypass Transition in a Turbulent Boundary Layer}

{DNS of bypass transition} in a flat-plate boundary layer is performed using synthetically generated free-stream turbulence (FST) as the inflow boundary condition, following the procedures of \citet{schlatter2001direct}, \citet{jacobs2001simulations}, and \citet{brandt2004transition}. The FST field is constructed using 3D wavenumber vectors located at the vertices of a dodecahedron. Each vector is rotated to ensure that the spanwise component ($\kappa_z$) satisfies periodic boundary conditions while preserving its original magnitude. For each wavenumber, {Orr–Sommerfeld and Squire modes are combined, with weighting factors chosen to ensure isotropy.
The resulting velocity fluctuations for each wavenumber are normalized to match the target turbulent kinetic energy corresponding to that wavenumber magnitude, as prescribed by the von Kármán energy spectrum in \citet{brandt2004transition}:} 
\[
E(\kappa) = \frac{2}{3} \kappa L_I \, 
\frac{1.606 (\kappa L_I)^4}{\left(1.350 + (\kappa L_I)^2 \right)^{17/6}} .
\]
\noindent  {Here, $\kappa$ denotes the wavenumber magnitude, and $L_I$ is the integral length scale.} The minimum and maximum wavenumber magnitudes are constrained by the spanwise domain size and the grid resolution. This synthetic FST serves as the inflow condition and initiates the formation of turbulent spots within the boundary layer.

\begin{figure*}[t]
    \centering
    \begin{subfigure}{0.8\textwidth}
        \includegraphics[width=\linewidth,trim=0 100 0 0, clip]{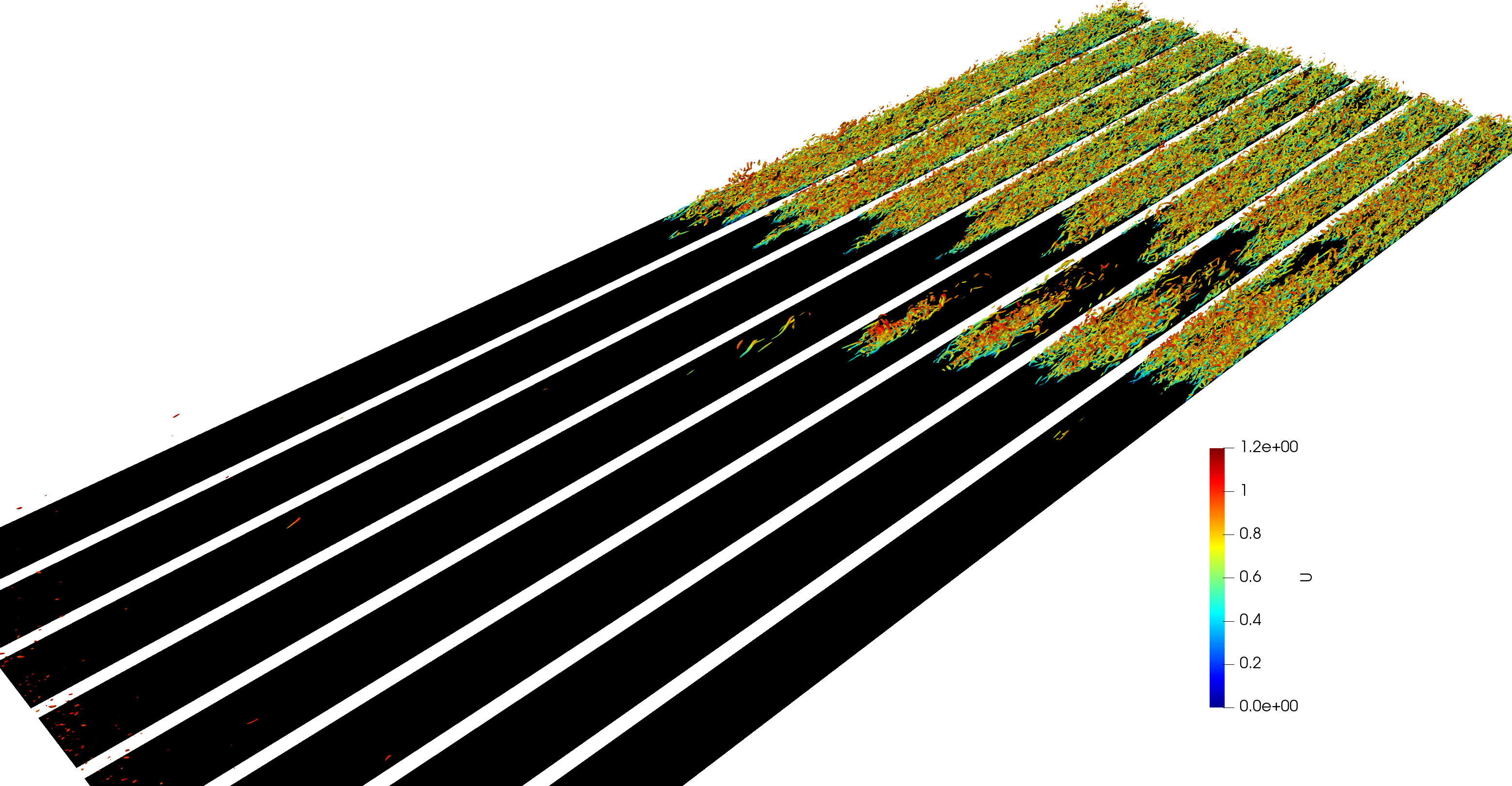}
        \label{fig:top}
        \vspace{-1.5em}
        \caption{}
    \end{subfigure}
    \par\vspace{0em}
    \centering
    \begin{subfigure}{0.37\textwidth}
        \centering
        \includegraphics[width=\linewidth,trim=0 0 0 0, clip]{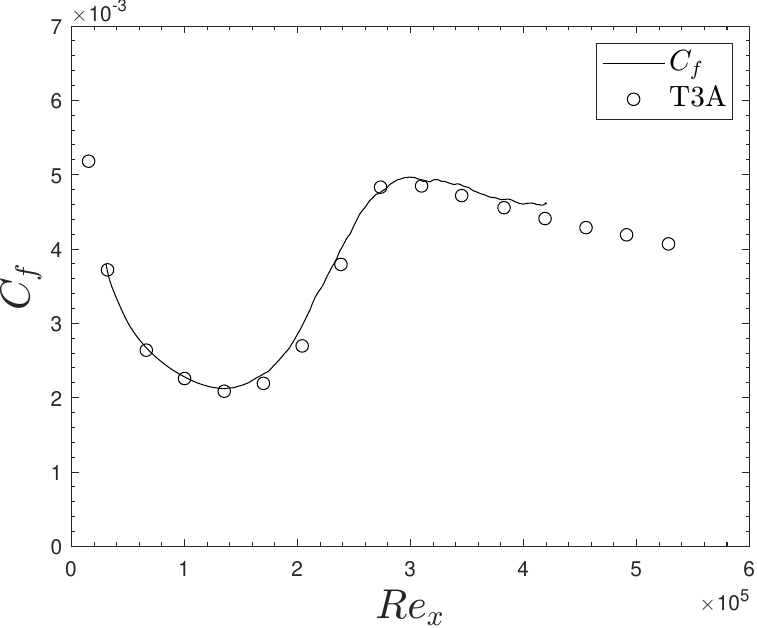}
        \label{fig:vortex1}
        \vspace{-1.5em}
        \caption{}
    \end{subfigure}
    \hspace{0.01\textwidth} 
    \begin{subfigure}{0.39\textwidth}
        \centering
        \includegraphics[width=\linewidth,trim=0 0 0 0, clip]{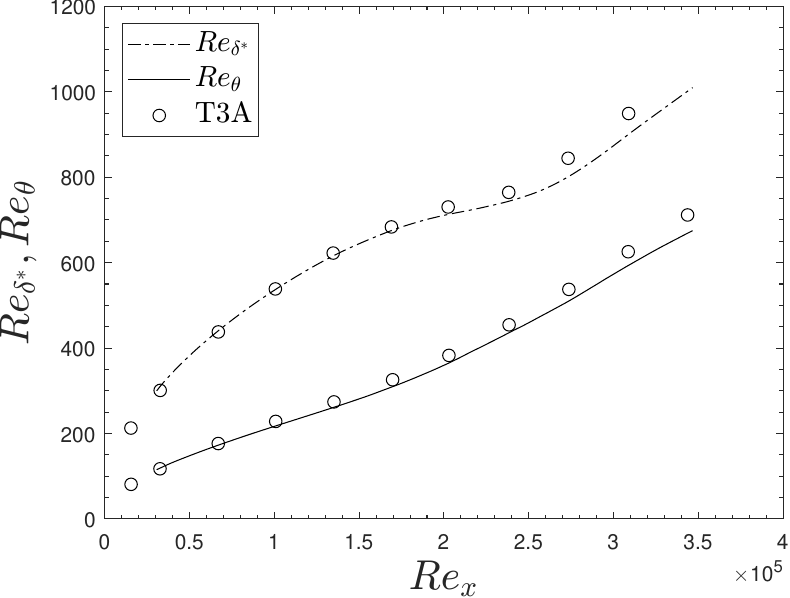}
        \label{fig:vortex2}
        \vspace{-1.5em}
        \caption{}
    \end{subfigure}
    \caption{{Bypass transition in boundary-layer flow: (a) time-sequence snapshots (top left to bottom right) of Q-criterion iso-surfaces showing the development of a turbulent spot, and comparisons with experimental T3A data for} (b) skin friction and (c) boundary-layer thickness parameters $\delta^*$ and $\theta$.}
    \label{fig:bypasstransition}
\end{figure*}

To reproduce the experimental conditions of the T3A case by \citet{roach1990influence}, {the inlet FST intensity is set to 3.6\%, and the integral length scale is prescribed as $L_I = 6$ in von karman energy spectrum. The simulation is performed at $Re_{\delta^*_0} = 300$ within a computational domain of $(L_x, L_y, L_z) = (1300, 60, 50)$, discretized using $(N_x, N_y, N_z) = (1536, 256, 192)$. Statistical averaging is conducted over 50{,}000 time steps, with each step requiring approximately 0.49 seconds.}
As shown in Fig.\ref{fig:bypasstransition}, the simulation results closely match the experimental data from the T3A case. Minor discrepancies remain, as also observed in previous DNS studies\citep{jacobs2001simulations, schlatter2001direct}, where perfect agreement was not achieved either.

\subsection{H-Type Natural Transition in Channel Flow}
DNS of H-type transition in channel flow is performed by introducing an initial disturbance field derived from the Orr–Sommerfeld equation. This fourth-order equation is solved using the Chebyshev collocation method to obtain accurate eigenvalues. A verification test confirms that the growth rate of a two-dimensional Orr–Sommerfeld mode matches theoretical predictions.
{Following the general setup} of \citet{lee2017simulations} at zero Weissenberg number to eliminate viscoelastic effects, the simulation initiates transition through the superposition of three eigenfunctions corresponding to wavenumber pairs {$(\alpha,\beta) = (1, 0)$, $(0.5, -1)$, and $(0.5, 1)$}, where $\alpha$ and $\beta$ denote streamwise and spanwise wavenumbers, respectively. The initial disturbance amplitudes are set to 0.01, 0.0002, and 0.0002 for the respective modes.

\begin{figure}[t]
   \begin{center}
    \begin{subfigure}[b]{0.45\textwidth}
        \centering
        \includegraphics[page=1, width=\linewidth,trim=200 50 950 50, clip]{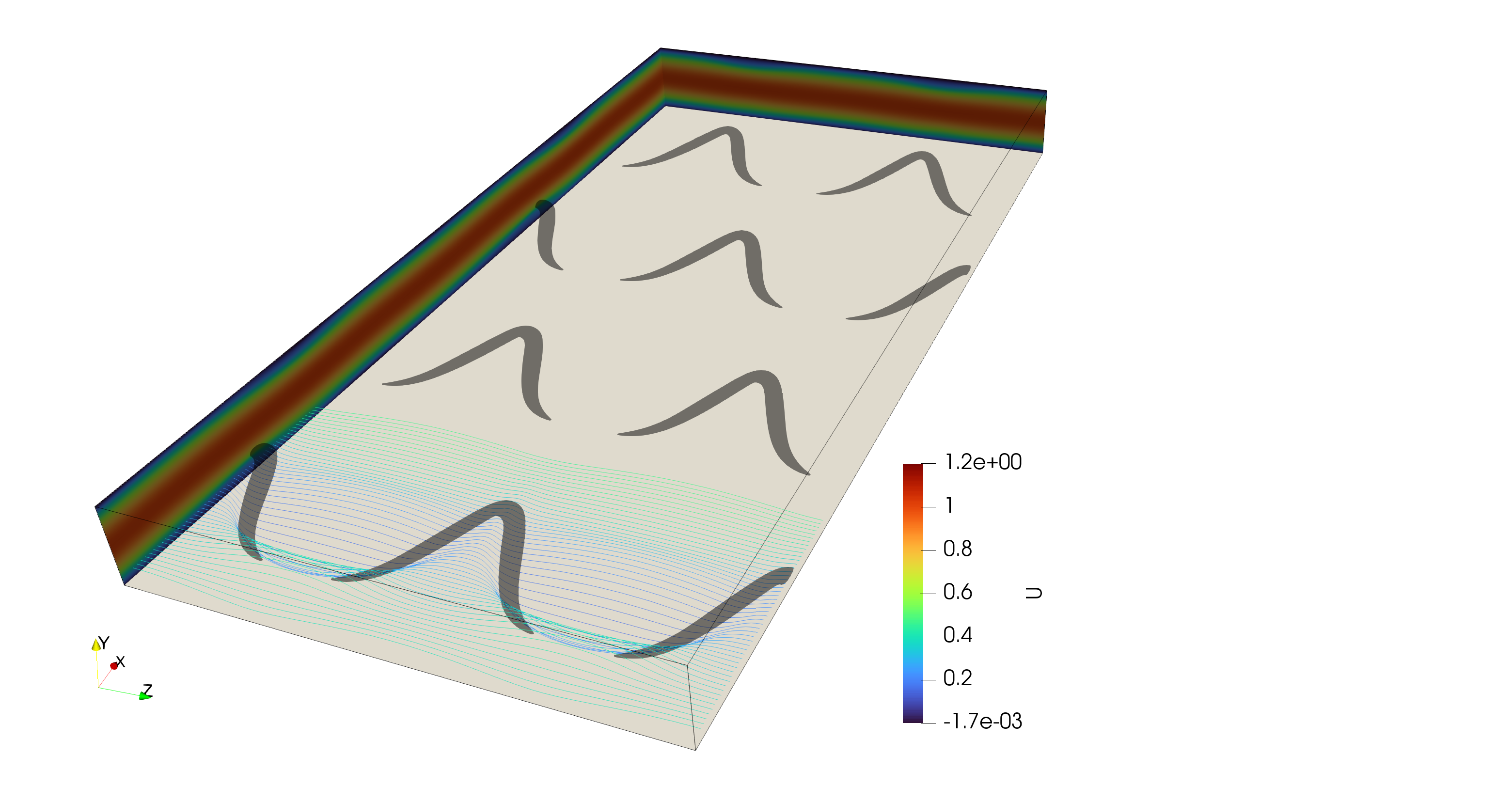}
        \caption{$\Lambda$ vortex }
        \label{fig:start}
    \end{subfigure}
    \hspace{0.01\textwidth}
    \begin{subfigure}[b]{0.45\textwidth}
        \centering
        \includegraphics[page=1, width=\linewidth,trim=200 50 950 50, clip]{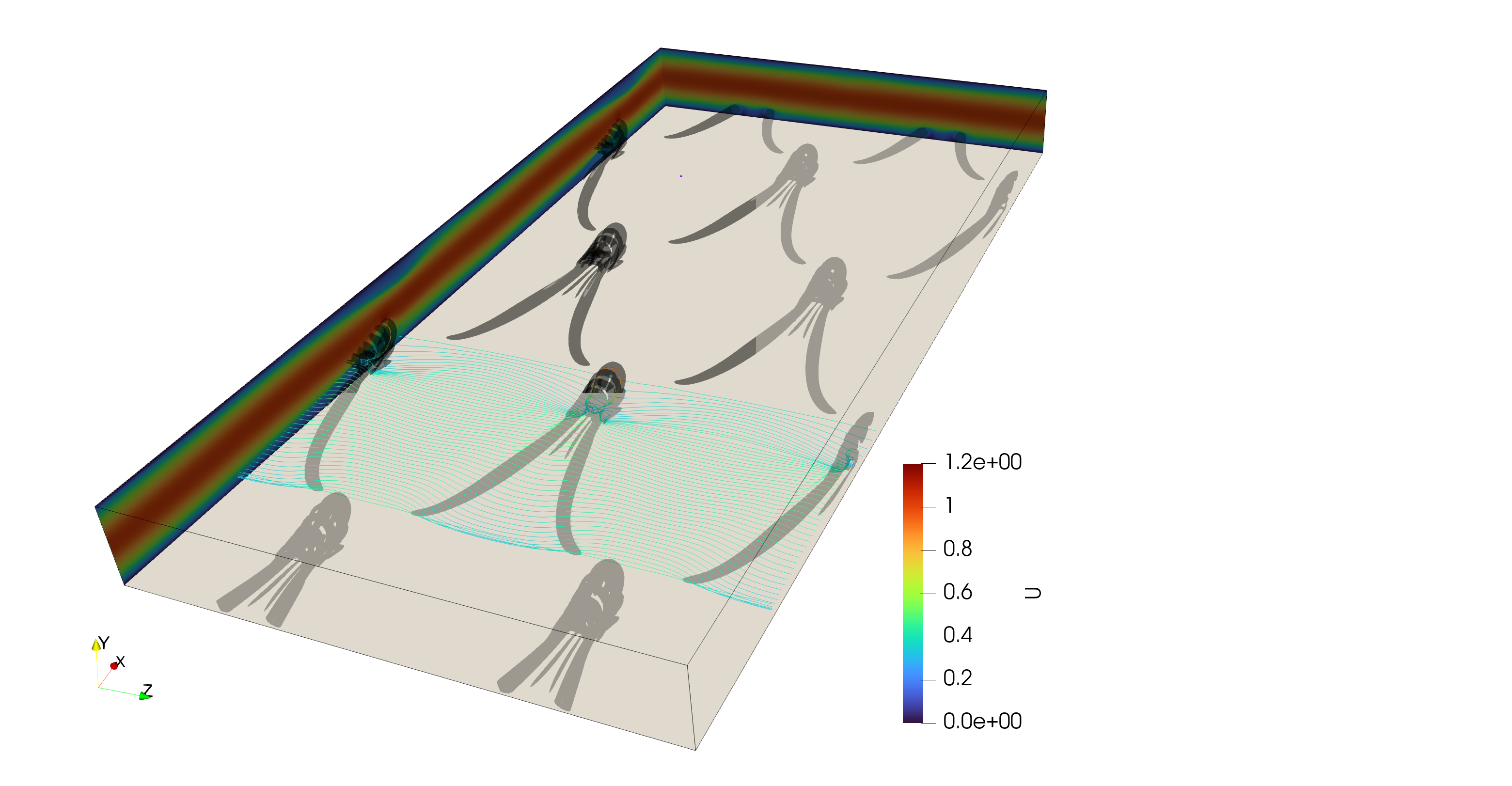}
        \caption{{Hairpin vortex}}
        \label{fig:end}
    \end{subfigure}
    \vfill
    \end{center}
    \caption{{Visualization of Q-criterion iso-surfaces in the lower half of the channel during H-type transition, with vortex lines colored by streamwise velocity $u$.}}
    \label{fig:Htypetransition}
\end{figure}

The simulation is conducted at a bulk Reynolds number of {$Re_b = 5333.3$} in a computational domain of $(L_x, L_y, L_z) = (8\pi h, 2h, 4\pi h)$, discretized with $(N_x, N_y, N_z) = (512, 512, 512)$. Each time step requires approximately 0.3 seconds on the target system.
As shown in Fig.~\ref{fig:Htypetransition}, the flow field exhibits the orderly formation of $\Lambda$ vortices and \textcolor{blue}{hairpin vortex}.

\section{Performance Analysis}

Performance evaluations were conducted on the JURECA supercomputing system at the Jülich Supercomputing Centre, utilizing 16 GPU nodes with a total of 64 NVIDIA A100 SXM GPUs. Each A100 GPU is equipped with 40GB of memory, yielding a total available GPU memory of 2.56TB. 
The code was compiled and executed using the NVIDIA HPC SDK (NVHPC 25.1) with CUDA 12 support and OpenMPI 5.0.5. CUDA-aware MPI communication was enabled via the MPI-settings/CUDA-UCC module.
In this study, three TDMA solvers and one Poisson solver were tested:

\begin{enumerate}
\item \textbf{PPT-v0}: divide-and-conquer TDMA (PPT method from~\citet{sun1989parallel})
\item \textbf{Pipelined-v1}: pipelining + divide-and-conquer TDMA (PPT-v0)
\item \textbf{Pipelined-v2}: pipelining + divide-and-conquer TDMA (PPT-v0 + even distribution strategy~\citep{kim2021pascal_tdma})
\item \textbf{Poisson-v1}: Pipelined-v2 + optimized FFT memory access (used only in full flow solver)
\end{enumerate}

\subsection{Test Configuration for the TDMA}

To isolate TDMA performance and accommodate larger problem sizes, the solver is benchmarked independently of the flow solver. The tests use 3D complex-valued data, with tridiagonal systems aligned along the wall-normal ($y$) direction, each of size $N_y$.

The domain size is carefully chosen to prevent excessive memory usage while preserving flexibility for batch size tuning. This design avoids constraints that would otherwise limit the choice of optimal batch sizes. Additionally, matrix sizes that are too small are excluded to ensure that modern GPUs can operate at full performance.
For strong scaling tests, the domain is fixed at $1024 \times 2048 \times 2048$. For weak scaling, the domain size scales with GPU count as $2048 \times (256 \times \text{GPUs}) \times 2048$.

Each batch consists of TDM systems that span the full $x$-direction and a subset of the $z$-direction. Each batched domain has a size of $N_x \times N_y \times batch_z$, producing $N_z / batch_z$ TDM systems of size $N_y \times N_x \times batch_z$. If $N_z$ is not divisible by $batch_z$, one additional system of size $N_y \times N_x \times \text{mod}(N_z, batch_z)$ is created.

The performance measurements include both the TDMA solver and associated overhead: assembling the TDM systems and reinserting the results into the 3D domain.
These preprocessing and postprocessing steps are essential, as the layout of the 3D domain does not match the layout of independent matrix problems. Each system uses the same coefficients as the Poisson equation in the flow solver to maintain realistic computational loads.

\begin{figure*}[t]
    \centering
    \begin{subfigure}[b]{0.48\textwidth} 
        \centering
        \includegraphics[page=5, width=\linewidth]{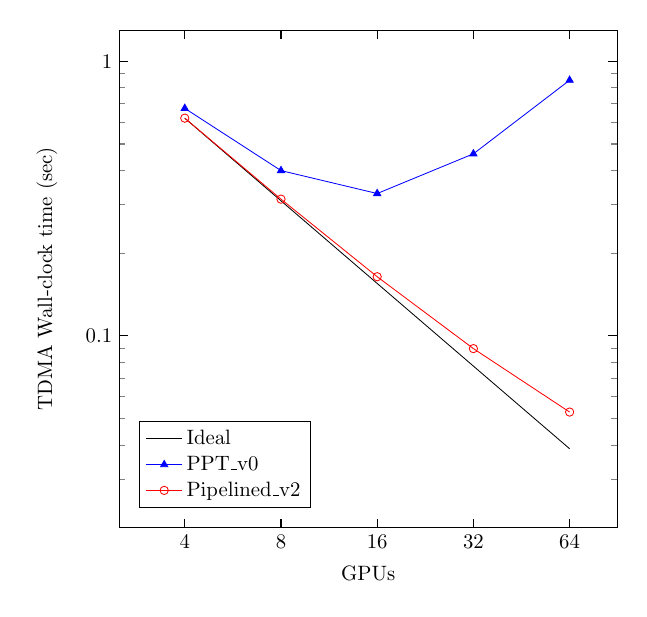}
        \caption{Strong scaling test}
        \label{fig:img1}
    \end{subfigure}
    \hspace{0.01\textwidth} 
    \begin{subfigure}[b]{0.48\textwidth} 
        \centering
        \includegraphics[page=7, width=\linewidth]{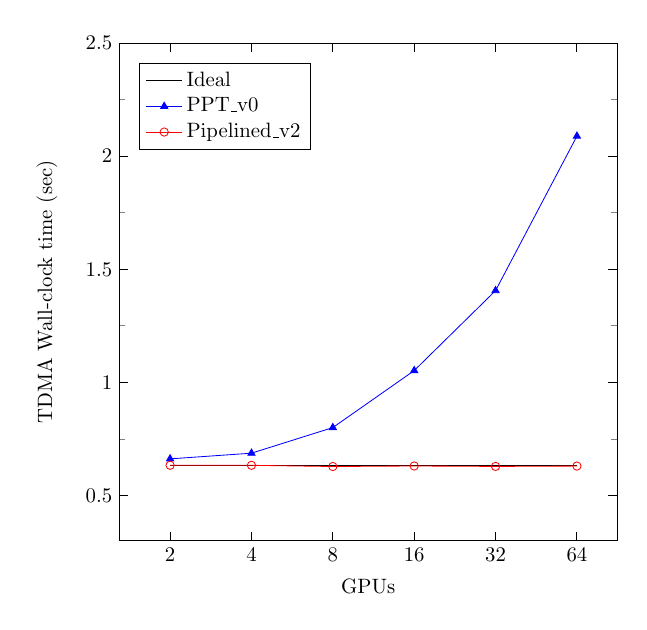}
        \caption{Weak scaling test}
        \label{fig:img2}
    \end{subfigure}
    \caption{
    Scalability of the ideal, original (PPT-v0), and enhanced (pipelined-v2) TDMA under two configurations: (a) Strong scaling test with a fixed domain size of $1024 \times 2048 \times 2048$. (b) Weak scaling test with a domain size scaled as $2048 \times (256 \times \text{GPUs}) \times 2048$ to maintain constant workload per GPU. Note that complex-number system are tested.}
    \vspace{0.1cm} 

\end{figure*}

\subsection{Strong Scalability Test of the TDMA Solver}

The strong scaling test was performed using a 3D domain of size $1024 \times 2048 \times 2048$. The TDMA solver was tested from 4 to 64 GPUs. For the 4-GPU configuration, the batch size was limited to 127 to prevent memory overflow. This constraint was not restrictive, as the optimal batch size in this case was 32. The maximum allowable batch size increased with GPU count: 128 for 8 GPUs, 256 for 16 GPUs, and 512 for 32 and 64 GPUs. As the number of GPUs increases, the memory per GPU decreases, allowing larger batch sizes. Scaling the batch size appropriately is essential to maintain sufficiently large TDM systems as the domain size per GPU decreases.

Figure~\ref{fig:img1} presents the strong scalability results. The proposed method, pipelined-v2, achieves 74.7\% scalability on 64 GPUs relative to the 4-GPU baseline. In contrast, the PPT-v0 shows only 4.9\% scalability and exhibits performance degradation beyond 16 GPUs.

Since the comparison is made between the raw PPT-v0 and the proposed method, the benefit of pipelining is not isolated, as the effect of the even distribution strategy by~\citet{kim2021pascal_tdma} is also present.  
To isolate the standalone effect of pipelining, the execution time of each stage (A, B, C, A2B+B2C) was analyzed to evaluate how much of the non-scalable stages (Stage~B, A2B, B2C) is hidden by pipelining.
To ensure accurate performance evaluation, measurements excluded the startup and termination phases of the pipeline. Each stage was timed independently by disabling all unrelated instructions. The average time for solving a single batched TDM system was also measured to evaluate the effectiveness of the pipelined algorithm by quantifying the hidden non-scalable stages' execution time.
Table~\ref{tab:results1} summarizes the execution time of each stage for pipelined-v1 and pipelined-v2 across different GPU counts. The first two cases used a batch size of 32, which was found to be optimal for pipelined-v2 at 4 GPUs. The remaining case used a batch size of 128 with pipelined-v2, which was optimal for 64 GPUs.
Key metrics include the original and actual impact of the non-scalable stages. The original impact denotes the total execution time of the non-scalable stages prior to pipelining. The actual impact is the residual cost after pipelining, computed by subtracting the scalable stages' execution time from the total execution time. Both metrics are expressed as percentages relative to the scalable stages' execution time.

\begin{table*}
\fontsize{9pt}{11pt}\selectfont
\captionsetup{width=\textwidth}
\caption{Strong sc
ling test: execution time of each stage and total execution time for solving single batched TDM systems across different GPU counts, using a fixed total domain size. The impacts of the non-scalable stage is also noted as a percentage based on the scalable stages' execution time
Results are shown for different TDMA versions (pipelined-v1 and pipelined-v2) and batch sizes (32 and 128).}
\begin{tabular*}{\linewidth}{@{\extracolsep{\fill}} cccccrccrr@{} } 
\toprule
\multirow{3}{*}{Type} & \multirow{3}{*}{GPUs$\;\;$} & \multirow{3}{*}{\textbf{Total}} & \multicolumn{3}{c}{Scalable Stages ($ms$)} & \multicolumn{4}{c}{Non-scalable Stages ($ms$)}  \\
\cmidrule(lr){4-6} \cmidrule(lr){7-10}
 & & & \;$A$ & $C$ & \textbf{Total\;\;} & $B$  & $A2B+B2C$ & \makecell{\textbf{Original}\\ \textbf{impact}} & {\scriptsize \makecell{\textbf{Actual impact}\\ after pipe- \\ lining ($ms$)}}\\
\midrule
\multirow{4}{*}{\shortstack[c]{Pipelined-v1\\$batch_z$ : 32}} 
 & $\;\;$8$\;\;$  & 5.26 $ms$ & 4.36 & 0.52 & 4.89 $ms$ & 0.46 & 0.79 & \textcolor{red}{26} $\%$\;\; & \textcolor{blue}{8} $\%$ $(0.37)$  \\
 & 16$\;\;$       & 3.26 $ms$ & 2.28 & 0.27 & 2.55 $ms$ & 0.85 & 1.49 & \textcolor{red}{92} $\%$\;\; & \textcolor{blue}{28} $\%$ $(0.71)$ \\
 & 32$\;\;$       & 3.62 $ms$ & 1.20 & 0.14 & 1.34 $ms$ & 1.60 & 3.12 & \textcolor{red}{352} $\%$\;\; & \textcolor{blue}{170} $\%$ $(2.28)$ \\
 & 64$\;\;$       & 7.08 $ms$ & 0.68 & 0.07 & 0.75 $ms$ & 3.04 & 6.14 & \textcolor{red}{1222} $\%$\;\; & \textcolor{blue}{842} $\%$ $(6.33)$ \\
\midrule
\multirow{4}{*}{\shortstack[c]{Pipelined-v2\\$batch_z$ : 32}}  
 & $\;\;$8$\;\;$  & 4.96 $ms$ & 4.37 & 0.52 & 4.90 $ms$ & 0.23 & 0.38 & \textcolor{red}{12} $\%$\;\; & \textcolor{blue}{1} $\%$ $(0.06)$ \\
 & 16$\;\;$       & 2.82 $ms$ & 2.28 & 0.27 & 2.55 $ms$ & 0.21 & 0.48 & \textcolor{red}{27} $\%$\;\; & \textcolor{blue}{11} $\%$ $(0.27)$ \\
 & 32$\;\;$       & 1.58 $ms$ & 1.20 & 0.15 & 1.35 $ms$ & 0.21 & 0.56 & \textcolor{red}{57} $\%$\;\; & \textcolor{blue}{18} $\%$ $(0.24)$ \\
 & 64$\;\;$       & 0.95 $ms$ & 0.68 & 0.08 & 0.76 $ms$ & 0.20 & 0.78 & \textcolor{red}{129} $\%$\;\;& \textcolor{blue}{25} $\%$ $(0.19)$ \\
\midrule
\multirow{4}{*}{\shortstack[c]{Pipelined-v2\\$batch_z$ : 128}} 
 & $\;\;$8$\;\;$  & 19.82 $ms$ & 17.67& 2.05 & 19.72 $ms$ & 0.48 & 1.09 & \textcolor{red}{8} $\%$\;\; & \textcolor{blue}{1} $\%$ $(0.11)$ \\
 & 16$\;\;$       & 10.08 $ms$ & 8.82 & 1.03 & 9.85 $ms$ & 0.35 & 1.26 & \textcolor{red}{16} $\%$\;\; & \textcolor{blue}{2} $\%$ $(0.23)$ \\
 & 32$\;\;$       & 5.57 $ms$ & 4.49 & 0.53 & 5.02 $ms$ & 0.29 & 1.35 & \textcolor{red}{33} $\%$\;\; & \textcolor{blue}{11} $\%$ $(0.56)$ \\
 & 64$\;\;$       & 3.27 $ms$ & 2.42 & 0.28 & 2.70 $ms$ & 0.30 & 1.70 & \textcolor{red}{74} $\%$\;\; & \textcolor{blue}{21} $\%$ $(0.57)$ \\
\bottomrule
\label{tab:results1}
\end{tabular*}
\end{table*}

\subsubsection{Pipelined-v1 with Batch Size 32}

Stage~A solves a TDM system of size $(2048 / p) \times batch_z \times 1024$, while Stage~C performs matrix operations {$4 \times (2048 / p) \times batch_z \times 1024$ FLOPs.} Their execution times decrease with increasing GPU count, consistent with the reduced per-GPU workload.
Stage~B solves a TDM system of size $2(p - 1) \times batch_z \times 1024$. As expected, its execution time increases linearly with GPU count, eventually surpassing the scalable stages' execution time after 32 GPUs. In this test, Stage~B handles sufficiently large systems and avoids low GPU occupancy, which makes overlapping with other computations ineffective.
Thus, pipelined-v1 contributes only by overlapping communication latency. 

During communication, Stage~A2B gathers $6p \times batch_z \times 1024$ data to a single GPU, and Stage~B2C scatters $2p \times batch_z \times 1024$ data back to all GPUs. Communication time therefore also increases linearly with GPU count.
At 8 and 16 GPUs, the measured non-scalable execution time closely matches that of Stage~B, indicating that communication latency (A2B + B2C) is almost fully hidden. However, at 32 and 64 GPUs, the communication cost surpasses that of Stage~B and cannot be entirely overlapped.

As a result, the total execution time of non-scalable stages becomes the dominant cost, reaching 1222\% of scalable stage time at 64 GPUs. Although much of the communication latency is hidden, the remaining overhead still accounts for 842\%, leading to overall performance degradation starting from 32 GPUs.

\subsubsection{Pipelined-v2 with Batch Size 32}

Pipelined-v2, which incorporates the method proposed by Kim et al.~\citep{kim2021pascal_tdma}, effectively mitigates the inefficiencies present in Pipelined-v1.
The configurations of Stage~A and Stage~C remain unchanged from Pipelined-v1. However, the matrix size in Stage~B is reduced from $2(p - 1) \times batch_z \times 1024$ in Pipelined-v1 to $2(p - 1) \times batch_z \times 1024/p$ in Pipelined-v2, resulting in a workload that remains small and nearly constant regardless of GPU count. 
However, as mentioned before, due to the extremely small TDM system size in Stage~B, this stage suffers from low performance caused by poor GPU occupancy.
This can be confirmed by comparing the execution times of Stages~A and B. On 8 GPUs, the matrix size decreases by approximately 146 times, while the execution time decreases by only 19 times.
Therefore, overlapping Stage~B with Stage~A and Stage~C effectively hides the performance loss associated with computing small TDM systems and enhances overall performance.

For communication, although the amount of data transferred in Stage~A2B and Stage~B2C remains the same as in Pipelined-v1, the improved communication pattern reduces overhead, preventing linear scaling with GPU count. Consequently, communication time becomes smaller than the computation time, allowing most of the communication latency to be hidden.

The results demonstrate substantial performance improvement. At 64 GPUs, the original non-scalable execution time accounted for 129\% of the scalable stages' execution time. With the proposed method, this figure is reduced to 25\%, significantly enhancing overall efficiency.
Thanks to the minimized non-scalable execution time, Stage~A emerges as the next bottleneck at 64 GPUs. Specifically, when scaling from 32 to 64 GPUs, Stage~A achieves only a 1.75× speedup instead of the ideal 2×. This limitation arises from the small size of each TDM system and is analyzed further in the next subsection using a larger batch size.

\subsubsection{Pipelined-v2 with Batch Size 128}

Testing Pipelined-v2 with a batch size of 128—identified as optimal for 64 GPUs—provides insight into mitigating non-scalable execution time at high GPU counts.
First, Stage~A shows improved performance due to the sufficiently large size of the TDM systems. Although the batch size increases by a factor of four, its execution time rises only 3.6×. Given that Stage~A dominates the total execution time, this improvement is significant.
The optimal batch sizes identified through the proposed strategy were 32 for 4 GPUs and 128 for 64 GPUs, validating its effectiveness.

Second, the relative share of communication time decreases as the batch size increases due to improved efficiency from larger message sizes, particularly when the problem size per GPU is small.
This observation may raise concerns that the pipelining strategy becomes less effective because the limited batch size required for scheduling multiple TDM systems may reduce communication efficiency.
However, this concern may be considered negligible, as communication performance does not continue to improve indefinitely with increasing batch size.
Benchmark studies show that the throughput of \texttt{MPI\_Alltoall} saturates beyond a modest message size, which is easily achieved with a moderate batch size for pipelining. On 128 NVIDIA A100 GPUs, performance scales nearly linearly once the message size per GPU exceeds 512KB\citep[see Fig.8(f)][]{chen2022highly}; this result is also clearly illustrated in the slides of\citep{Subramoni2024Boosting}. Similar behavior is observed on 64 NVIDIA V100 GPUs using OpenMPI 4.1.4 with UCX 1.12.1, where nearly linear scaling is reported between 512KB and 1MB~\citep[see Fig.~11(d)]{shafie2022high}.
In the present configuration, with a batch size of 128, the communication volumes per GPU reach approximately 2MB in Stage A2B and 6MB in Stage B2C, calculated as $(2 \times batch_z \times 1024) \times 8$ and $(6 \times batch_z \times 1024) \times 8$ bytes, respectively. These values exceed the throughput saturation thresholds, indicating that communication efficiency remains stable beyond this batch size.
Since the saturation threshold is already exceeded at the optimal batch size used in the 64-GPU strong-scaling tests—which involve quite small domain size for each GPU—communication inefficiency in \texttt{MPI\_Alltoall} is unlikely to pose a performance concern in most scenarios.

Another important observation is that some communication latency remains exposed, even when it is shorter than computation time. This phenomenon is clearly observed at 32 and 64 GPUs, where the actual non-scalable execution time after overlapping exceeds the Stage~B execution time. In these cases, unhidden communication overhead must exist to explain the non-scalable execution time exceeding the execution time of Stage~B.

Additionally, improving Stage~B performance offers limited benefit. When increasing the batch size from 32 to 128, the Stage~B matrix size grows fourfold, but execution time increases by only 1.5× due to better GPU occupancy. However, since Stage~B is overlapped with other computations and the performance loss due to small TDM system is hidden, low occupancy impact is not expected to meaningfully affect the total execution time.

Overall, larger batch sizes reduce the execution time of scalable stages by generating sufficiently large TDM systems in Stage~A and also diminish the impact of non-scalable stages.
For example, the impact of non-scalable execution time at 32 GPUs decreases from 18\% with a batch size of 32 to 11\% with a batch size of 128. 
This improvement is likely due to reduced communication time.
However, this approach introduces trade-offs: at the end of the pipeline, communication overhead cannot be fully hidden, and the penalty becomes more pronounced at excessively large batch sizes.

\subsection{Weak Scalability Test of the TDMA Solver}

The 3D domain for the weak scaling test was configured as $2048 \times (256 \times \text{GPUs}) \times 2048$, ensuring a fixed domain size per GPU. The TDMA solver was evaluated using 2 to 64 GPUs. To prevent memory overflow, the maximum batch size was limited to 128. This upper bound proved sufficient, as the optimal batch size across all configurations ranged between 21 and 64.

\begin{table*}
\fontsize{9pt}{11pt}\selectfont
\captionsetup{width=\textwidth}
\caption{Weak scaling test: execution time of each stage and total execution time for solving single batched TDM systems across different GPU counts, with a fixed domain size per GPU. The impacts of the non-scalable stage is also noted as a percentage based on the scalable stages' execution time. Results are shown for different TDMA versions (pipelined-v1 and pipelined-v2) with constant batch size.}
\begin{tabular*}{\linewidth}{@{\extracolsep{\fill}} cccccrccrr@{} } 
\toprule
\multirow{3}{*}{Type} & \multirow{3}{*}{GPUs$\;\;$} & \multirow{3}{*}{\textbf{Total}} & \multicolumn{3}{c}{Scalable Stages ($ms$)} & \multicolumn{4}{c}{Non-scalable Stages ($ms$)}  \\
\cmidrule(lr){4-6} \cmidrule(lr){7-10}
 & & & \;$A$ & $C$ & \textbf{Total\;\;} & $B$  & $A2B+B2C$ & \makecell{\textbf{Original}\\ \textbf{impact}} & {\scriptsize \makecell{\textbf{Actual impact}\\ after pipe- \\ lining ($ms$)}}\\
\midrule
\multirow{5}{*}{\shortstack[c]{Pipelined-v1\\$batch_z$ : 32}} 
 & $\;\;$4$\;\;$  & \!10.11 $ms$ & \;8.70 & 1.03 & 9.73 $ms$ & \;0.59 & 0.23 & \textcolor{red}{8} $\%$\;\; & \textcolor{blue}{4} $\%$ $(0.38)$ \\
 & $\;\;$8$\;\;$  & \!10.64 $ms$ & \;8.72 & 1.03 & 9.75 $ms$ & \;0.88 & 1.48 & \textcolor{red}{24} $\%$\;\; & \textcolor{blue}{9} $\%$ $(0.89)$ \\
 & 16$\;\;$       & \!12.14 $ms$ & \;8.71 & 1.02 & 9.73 $ms$ & \;1.67 & 2.86 & \textcolor{red}{47} $\%$\;\; & \textcolor{blue}{25} $\%$ $(2.41)$ \\
 & 32$\;\;$       & \!13.18 $ms$ & \;8.72 & 1.02 & 9.74 $ms$ & \;3.21 & 6.11 & \textcolor{red}{96} $\%$\;\; & \textcolor{blue}{35} $\%$ $(3.45)$ \\
 & 64$\;\;$       & \!16.38 $ms$ & \;8.70 & 1.02 & 9.72 $ms$ & \;6.11 & \!\!12.86 & \textcolor{red}{195} $\%$\;\; & \textcolor{blue}{68} $\%$ $(6.65)$ \\
\midrule
\multirow{5}{*}{\shortstack[c]{Pipelined-v2\\$batch_z$ : 32}}  
 & $\;\;$4$\;\;$  & \!9.82 $ms$ & \;8.71 & 1.02 & 9.73 $ms$ & \;0.39 & 0.57 & \textcolor{red}{6} $\%$\;\; & \textcolor{blue}{1.0} $\%$ $(0.09)$ \\
 & $\;\;$8$\;\;$  & \!9.82 $ms$ & \;8.73 & 1.02 & 9.75 $ms$ & \;0.27 & 0.85 & \textcolor{red}{9} $\%$\;\; & \textcolor{blue}{0.7} $\%$ $(0.07)$ \\
 & 16$\;\;$       & \!9.79 $ms$ & \;8.69 & 1.03 & 9.71 $ms$ & \;0.25 & 0.94 & \textcolor{red}{10} $\%$\;\; & \textcolor{blue}{0.8} $\%$ $(0.07)$ \\
 & 32$\;\;$       & \!9.79 $ms$ & \;8.68 & 1.02 & 9.71 $ms$ & \;0.24 & 1.09 & \textcolor{red}{11} $\%$\;\; & \textcolor{blue}{0.9} $\%$ $(0.08)$ \\
 & 64$\;\;$       & \!9.83 $ms$ & \;8.69 & 1.02 & 9.71 $ms$ & \;0.24 & 1.40 & \textcolor{red}{14} $\%$\;\; & \textcolor{blue}{1.2} $\%$ $(0.12)$ \\
\bottomrule
\label{tab:results2}
\end{tabular*}
\end{table*}

Figure~\ref{fig:img2} presents the results of the weak scaling test. The proposed method, pipelined-v2, achieves 100.5\% scalability on 64 GPUs relative to its baseline performance on 2 GPUs, while the PPT-v0 achieves only 31.7\%. The following subsections analyze how pipelined-v2 achieves ideal scalability. Table~\ref{tab:results2} reports the average execution time per stage for solving a single batched TDM system.

As in the strong scaling test, the execution time of each stage (A, B, C, A2B+B2C) and average time for solving a single batched TDM system was measured and shown in Table~\ref{tab:results2}. 

\subsubsection{Pipelined version 1 with batch size 32}

Stage~A solves a TDM system of size $256 \times batch_z \times 2048$, and Stage~C performs matrix operations totaling {$4 \times256 \times batch_z \times 2048$ FLOPs}. As a result, the execution times for Stage~A and Stage~C remain consistent across GPU counts.
Stage~B solves a TDM system of size $2(p - 1) \times batch_z \times 2048$, so its execution time increases proportionally with GPU count.

The communication time for Stages A2B and B2C also increases approximately linearly with GPU count. 
Despite this, nearly all communication latency is overlapped with computation, as the communication latency remains smaller than the total computation time. As a result, total execution time closely matches the sum of computation stages. 
Nonetheless, ideal scalability is not achieved due to the gradual increase in Stage~B's computation time.
An anomaly is observed at the 4-GPU configuration, where Stage~A2B communication time appears unusually low, possibly because all GPUs reside on the same node.

Overall, the behavior is similar to the strong scalability results for pipelined-v1. However, due to the relatively small contribution of non-scalable stages, the total execution time does not increase significantly.

\subsubsection{Pipelined Version 2 with Batch Size 32}

As shown in the strong scaling test analysis, the size of the TDM system in Stage~B decreases from $2(p - 1) \times batch_z \times 2048$ to $2(p - 1) \times batch_z \times 2048/p$, resulting in nearly constant computation time. 
Since Stages A and C also maintain constant execution times in the weak scaling test, the constant execution time of Stage~B does not negatively affect overall scalability.
Additionally, the relative share of communication latency remains minimal. As communication occupies only a small fraction of the total execution time, its latency is effectively hidden by pipelining.

Therefore, at 64 GPUs, the non-scalable stages' execution time accounts for only 1.2\% (down from 14\%) of the scalable stages' execution time, indicating that weak scaling performance remains nearly ideal and is likely to remain stable even beyond 64 GPUs.
This high scalability arises from two key factors: Stage~B does not negatively affect overall scalability and the sufficiently large domain size allocated to each GPU. In contrast, strong scaling performance becomes limited in the current test setting, where each GPU handles a smaller subdomain to allow execution of the same problem across only 4 GPUs.

\subsection{Summary of the Impact of Pipelining for TDMA}

As intended, most of the communication latency and the performance degradation caused by small TDM systems in Stage~B are effectively hidden within the computation stages.
Because communication latency (Stage~A2B and B2C) and Stage~B constitute the entirety of the non-scalable stages, the overall non-scalable execution time is substantially reduced.
To provide a more meaningful comparison—rather than simply contrasting with raw PPT-v0—the version only without pipelining is compared against the pipelined version.
Even at the largest GPU count of 64, the non-scalable stage accounts for only 21\% of the scalable stage time in strong scaling tests (down from 74\%), and 1.2\% in weak scaling tests (down from 14\%), excluding the start and end phases of the pipeline.

These results confirm the substantial performance improvement achieved by the pipelined-v2 method over not only the original PPT, but also the baseline TDMA without pipelining (PPT-v0 combined with the even-distribution scheme~\citep{kim2021pascal_tdma}). 
The proposed pipelined structure consistently hides a significant portion of non-scalable execution time across varying GPU counts and domain sizes, offering clear performance advantages even at large GPU scales.
This structure is expected to remain effective as long as the communication time stays below the computation time—a condition that held even for very small per-GPU domains in the 64-GPU strong scaling tests. Therefore, the approach appears broadly applicable to most practical cases.
This substantial reduction in non-scalable stages enables the development of a highly scalable TDMA algorithm that is well suited for large-scale multi-GPU environments.

\subsection{Flow Solver Test}

\begin{table}
\fontsize{10pt}{13pt}\selectfont
\captionsetup{width=\textwidth}
\caption{Parallel efficiency and performance improvement of flow solver components with the enhanced TDMA and Poisson solvers. The comparison is done by evaluating each enhanced flow solver component against its counterpart in the flow solver of~\citet{ha2021multi}. The test uses a domain size of $1024 \times 128 \times 1024$ per GPU, with parallel efficiency measured relative to the 2-GPU baseline. The N/A region indicates redundant data, as the results of \textbf{Poisson-v2} are identical to those of \textbf{Pipelined-v2} in these cases.}
\label{tbl1}

\centering
\begin{tabular*}{0.7\textwidth}{@{\extracolsep{\fill}} clrrr@{} }

        \toprule
    \multirow{2}{*}{GPUs} & \multirow{2}{*}{Target} & \multicolumn{1}{c}{Parallel efficiency} & \multicolumn{2}{c}{Performance improvement}  \\
    \cmidrule(lr){3-3} \cmidrule(lr){4-5}
     & & \multicolumn{1}{c}{Pipelined-v2} & \multicolumn{1}{c}{Pipelined-v2} & \multicolumn{1}{c}{Poisson-v1} \\
        \midrule
        \multirow{3}{*}{4} & TDMA solver & $0.71 \;\Rightarrow\; 1.01$ \;& $20\%$ \;& $N\!/\!A$ \;     \\
                           & Poisson solver & $0.85 \;\Rightarrow\; 1.01$  \;& $9\%$ \;& $24\%$ \;\\
                           & Flow solver & $0.98 \;\Rightarrow\; 1.00$  \;& $2\%$  \;& $6\%$  \;\\
        \hline
        \multirow{3}{*}{8} & TDMA & $0.61 \;\Rightarrow\; 0.99$ \;& $37\%$ \;& $N\!/\!A$    \;\\
                           & Poisson & $0.79 \;\Rightarrow\; 1.00$ \;& $18\%$ \;& $33\%$ \;\\
                           & Flow & $0.93 \;\Rightarrow\; 0.97$  \;& $4\%$  \;& $7\%$ \;\\
        \hline
        \multirow{3}{*}{16} & TDMA & $0.47 \;\Rightarrow\; 0.98$ \;& $83\%$ \;& $N\!/\!A$    \;\\
                           & Poisson & $0.67 \;\Rightarrow\; 0.99$ \;& $39\%$ \;& $58\%$ \;\\
                           & Flow & $0.46 \;\Rightarrow\; 0.96$  \;& $9\%$  \;& $11\%$ \;\\
        \hline
        \multirow{3}{*}{32} & TDMA & $0.30 \;\Rightarrow\; 0.97$ \;& $171\%$ \;& $N\!/\!A$    \;\\
                           & Poisson & $0.51 \;\Rightarrow\; 0.99$ \;& $81\%$ \;& $105\%$ \;\\
                           & Flow & $0.77 \;\Rightarrow\; 0.90$  \;& $17\%$  \;& $20\%$ \;\\
        \hline
        \multirow{3}{*}{64} & TDMA & $0.19 \;\Rightarrow\; 0.96$ \;& $337\%$ \;& $N\!/\!A$    \;\\
                           & Poisson & $0.35 \;\Rightarrow\; 0.98$ \;& $161\%$ \;& $195\%$ \;\\
                           & Flow & $0.63 \;\Rightarrow\; 0.81$  \;& $31\%$  \;& $33\%$ \;\\
                           \bottomrule
        \Xhline{1\arrayrulewidth}
\label{tab:results3}
\end{tabular*}
\end{table}

This section evaluates the impact of the TDMA and Poisson solvers on each component of the flow solver, in comparison with the baseline implementation of \citet{ha2021multi}, as summarized in Table~\ref{tab:results3}. The grid size is fixed at 0.163 billion cells per GPU. The enhancements of momentum equation is not included in this test.

First, the pipelined TDMA achieves near-ideal weak scaling up to 64 GPUs, whereas the original solver achieves only 18.7\% parallel efficiency relative to the 2-GPU baseline. This improvement increases the overall parallel efficiency of the flow solver from 63\% to 81\% on 64 GPUs and shifts the Poisson solver from being the primary bottleneck to the most scalable component.  Enhancements to the momentum equation are excluded from this evaluation.

The speedup of the standalone TDMA solver increases significantly with GPU count, reaching a 337\% improvement at 64 GPUs. In contrast, the speedup for the full flow solver is 31\% under the same conditions. The computational workload of the TDMA portion within the Poisson solver drops from 30.5\% to 9.1\% in the weak scaling test. 
When Poisson-v1 is adopted, FFT operations within the solver are further optimized, resulting in an additional performance gain of about 13\% compared to the baseline Poisson solver in most test cases.

\subsection{Discussion about 1D/2D domain decomposition}

\begin{table*}
\fontsize{9pt}{11pt}\selectfont
\captionsetup{width=\textwidth}
\caption{Comparison of domain decomposition strategies for FFT-based Poisson solvers. The table contrasts 1D slab decomposition—employing either a divide-and-conquer approach or global transpose for solving TDM systems—with 2D pencil decomposition, which applies the divide-and-conquer method in the $y$-direction and a global transpose in the $x$-direction.}
\begin{tabular*}{\linewidth}{@{\extracolsep{\fill}} c c!{\vrule width 0.3pt} ccc@{}}
\toprule
\multicolumn{2}{c!{\vrule width 0.3pt}}{} & strategy A & strategy B & strategy C \\
\cmidrule(lr){3-5}
\multicolumn{2}{c!{\vrule width 0.3pt}}{Decomposition type} & 1D($y$) slab & 1D($y$) slab & 2D($xy$) pencil \\ [1ex]
\cmidrule(lr){3-5}
\multicolumn{2}{c!{\vrule width 0.3pt}}{TDMA strategy ($y$-direction)} & Divide-and-conquer & Global transpose & Divide-and-conquer \\
\multicolumn{2}{c!{\vrule width 0.3pt}}{FFT strategy \hspace{10pt}($x$-direction)} & -- & -- & Global transpose \\
\midrule \\ [-2.5ex]
\multirow{2}{*}{\shortstack[c]{\hspace{30pt}Number of partitions}} 
& {\vrule width 0.3pt}\;$y$-direction & $p$ & $p$ & $p_y \;(p/2,p/4,p/8)$ \\
& {\vrule width 0.3pt}\;$x$-direction & $1$ & $1$ & $p_x \;(2,4,8)$ \\[0.2ex]
\cmidrule(lr){1-5}
\multirow{2}{*}{\shortstack[c]{\hspace{30pt}Total data volume for \\ \hspace{30pt}comunication ($\times N_xN_z$)}} 
& {\vrule width 0.3pt}\;$y$-direction & $8p$ & $2N_y$ & $8p_y$ \\
& {\vrule width 0.3pt}\;$x$-direction & -- & -- & $2N_y$ \\
\cmidrule(lr){1-5}
\multirow{2}{*}{\shortstack[c]{\hspace{30pt}Number of GPUs per\\ \hspace{30pt}\texttt{MPI\_Alltoall}}} 
& {\vrule width 0.3pt}\;$y$-direction & $p$ & $p$ & $p_y$ \\
& {\vrule width 0.3pt}\;$x$-direction & -- & -- & $p_x$ \\[0.2ex]
\cmidrule(lr){1-5}
\multirow{2}{*}{\shortstack[c]{\hspace{30pt}Communication scope}} 
& {\vrule width 0.3pt}\;$y$-direction & inter-node & inter-node & inter-node \\
& {\vrule width 0.3pt}\;$x$-direction & -- & -- & intra-node \\[0.2ex]
\cmidrule(lr){1-5}
\multicolumn{2}{c!{\vrule width 0.3pt}}{Relative communication cost (theoretical estimate)} & $36p$ & $9N_y$ & $36p_y + N_ylog_2(p_x)$ \\
\cmidrule(lr){1-5}
\multicolumn{2}{c!{\vrule width 0.3pt}}{Performance threshold for outperforming Strategy A} & -- & $\textcolor{red}{N_y/p {<} 4}$ & $\textcolor{red}{N_y/p {<} 18}\;(p_x=2)$ \\
\cmidrule(lr){1-5}
\multicolumn{2}{c!{\vrule width 0.3pt}}{Ghost cell size per GPU ($\times N_z$)} & $2pN_x$ & $2pN_x$ & $2p_xN_y + 2p_yN_x$ \\
\bottomrule
\label{tab:results4}
\end{tabular*}
\end{table*}

\subsubsection{Advantages and Limitations of 1D Slab-Type Decomposition}
As shown in the previous section, the flow solver maintains strong parallel efficiency under 1D slab-type decomposition up to 64 GPUs. Therefore, for moderately large-scale simulations, 1D decomposition remains an attractive option. It simplifies implementation and significantly reduces post-processing overhead, particularly in canonical boundary-layer and channel flows. These simulations often rely on spectral analysis in both the streamwise and spanwise directions~\citep{lee2015direct, wu2012direct, del2009estimation}. Retaining these directions undecomposed allows FFT-based post-processing to be performed entirely within a single GPU, thereby eliminating the need for global all-to-all communication.

However, as the number of GPUs increases, the number of grid cells per GPU in the wall-normal ($y$) direction must decrease to maintain uniform resolution across all spatial directions. When the $y$-direction grid resolution per GPU becomes too small, the primary advantage of 1D decomposition—avoiding all-to-all communication—deteriorates rapidly. For instance, the extreme-scale DNS at $Re_\lambda = 2500$ by \citet{yeung2025gpu} employed 35 trillion grid cells across 32,768 GPUs, leading to $N_y/p \approx 1$, which renders 1D slab decomposition impractical. Under such conditions, transitioning to 2D pencil-type decomposition becomes unavoidable.

\subsubsection{Comparison of domain decomposition strategies (1D vs 2D)}
The communication costs of three decomposition strategies are compared: 1D slab with divide-and-conquer TDMA (Strategy A), 1D slab with global transpose (Strategy B), and 2D pencil decomposition (Strategy C). To evaluate each approach for solving the Poisson equation, Table~\ref{tab:results4} presents theoretical communication metrics. Although ghost cells are not directly involved in the Poisson solver, they are included because they significantly impact the communication overhead of broader numerical schemes such as finite-difference or finite-volume methods.

Consider an FFT-based Poisson solver where the TDM systems are aligned along the $y$ direction. The 1D slab decomposition aligns with the $y$ direction, while the 2D pencil decomposition partitions the domain along the $x$ direction. Let $N_x$, $N_y$, and $N_z$ denote the number of grid cells in the $x$-, $y$-, and $z$-directions, respectively, and $p$ the total number of GPUs.

First, consider Strategy A (1D slab with divide-and-conquer TDMA) and Strategy B (1D slab with global transpose). As summarized in Table~\ref{tab:results4}, the divide-and-conquer approach incurs a total communication volume of $8p$ per TDM system—$6p$ for Stage A2B and $2p$ for Stage B2C—using \texttt{MPI\_Alltoall}. In contrast, Strategy B requires two global all-to-all communications, resulting in a total volume of $2N_y$. When $N_y/p \leq 4$, the communication overhead of the divide-and-conquer approach becomes comparable to or exceeds that of global transpose. Conversely, for $N_y/p \geq 8$, the divide-and-conquer method exhibits superior communication efficiency.

Next, Strategy A is compared with Strategy C (2D pencil-type decomposition). The most efficient configuration applies a small number of partitions (typically 2–8) along one direction to enable intra-node communication, while the other direction adopts finer partitioning. This strategy was employed by \citet{sanhueza2025pencil} for finite-difference methods and by \citet{yeung2025gpu} for {pseudo-spectral methods}.
For finite-difference methods, fast intra-node communication is particularly advantageous in the $x$ direction, where heavy all-to-all communication associated with global transposes occurs. Consequently, finer partitions along the $y$ direction—used for divide-and-conquer TDMA—are adopted, as in \citet{sanhueza2025pencil}. 
Communication cost under this configuration is estimated using the results of \citet{yeung2025gpu}, who used 32,768 AMD MI250X GPUs interconnected without CPU involvement. In their study, \texttt{MPI\_Alltoall} latency scales with the total data volume and $\log(p)$, as shown in the R-MPI cases in Tables 3 and 5. For inter-node communication in the $y$ direction, latency remains roughly constant with respect to partition count and is primarily influenced by data volume, as shown in the C-MPI cases.
Additionally, inter-node communication was observed to be roughly three times more expensive than intra-node communication when using 8 GPUs per node ($p_r = p_c = 8$). {Based on this communication time estimation}, 2D pencil decomposition outperforms 1D slab decomposition when {$N_y/p < 18$ with $p_x = 2$, $N_y/p < 13.5$ with $p_x = 4$, and $N_y/p < 10.5$ with $p_x = 8$}. Beyond communication costs, 1D slab decomposition also generates significantly larger halo (ghost) regions compared to 2D pencil decomposition~\citep{van2015pencil}, further increasing overhead in general numerical schemes beyond the Poisson solver.

\subsubsection{Usefulness of the Proposed TDMA in 2D Pencil-Type Decomposition}
As discussed earlier, even under 2D pencil decomposition, $p_x$ is typically kept small to preserve intra-node communication. Consequently, finer partitioning—assigning a larger number of GPUs along the TDMA direction—remains essential, making the scalability of TDMA a critical factor. Additionally, by adjusting the decomposition—specifically, increasing $p_y$ while reducing $p_x$ (e.g., from 8 to 4 or from 4 to 2)—{a large portion of the communication burden} can be shifted to the TDMA stage, thereby reducing the all-to-all communication cost associated with FFT operations. {This shift is particularly advantageous because TDMA entails substantially higher computational cost than FFT; in a single-GPU A100 test on a $1024 \times 128 \times 1024$ grid, FFT + IFFT in one direction required only 34.0\% of the TDMA execution time. Such long computational time allows communication overhead to be effectively hidden through computation–communication overlap.}

Finally, apart from the Poisson solver, the momentum equation involves only the solution of tridiagonal systems and does not require FFTs. As a result, its performance remains largely unaffected by the choice between 1D and 2D decomposition.

\subsection{Comparison of Poisson Solver with Recent Work}

The present Poisson solver is compared with that of \citet{sanhueza2025pencil}, using 64 NVIDIA A100 GPUs and similar grid sizes.  
Since the structure of the full flow solver may differ, hindering a direct comparison of TDMA performance, the Poisson solver alone is used as the basis for evaluation.

Although this comparison provides meaningful insight, several differences between the two implementations should be noted.
First, the two solvers adopt different base divide-and-conquer methods. \citet{sanhueza2025pencil} implements the approach of \citet{laszlo2016manycore}, incorporating the even-distribution strategy of \citet{kim2021pascal_tdma}. In contrast, the current solver builds on the divide-and-conquer method of \citet{sun1989parallel}, also applying the even-distribution strategy, and further enhances scalability through the proposed pipelined algorithm. Despite these differences, both solvers share a similar structural foundation, and the execution time is not expected to differ significantly due to the choice of base method alone.

Second, the FFT procedures used in the two studies differ. The current implementation performs a half-cosine transform in one direction instead of an FFT to handle the non-periodic spanwise direction in boundary layer simulations, whereas \citet{sanhueza2025pencil} applies two full FFTs for channel flow. Since the half-cosine transform incurs additional overhead—effectively requiring two FFTs—adjustments were made to allow for a fairer performance comparison.
In a benchmark case with a $1024 \times 8192 \times 1024$ grid, the present Poisson solver required 98.62 ms, composed of 53.65 ms for the TDMA and 44.97 ms for FFT and inverse FFT. The ratio of execution time between a half-cosine transform + FFT and FFT + FFT is measured as 1:0.624 using a Tesla P100. Although this measurement was taken on a different GPU and domain, the difference in transform count (3 FFTs versus 2) supports the assumption that the ratio remains approximately valid. Accordingly, the FFT portion is scaled from 44.97 ms to 28.04 ms for comparison.

After this adjustment, the total execution time for the Poisson solver in the present study is 81.69 ms for 85.9 billion grid cells, compared to 114.9 ms reported by \citet{sanhueza2025pencil} for 99.5 billion grid cells. (The latter value was estimated from a graph and may contain minimal error.)
While the current solver exhibits slightly better performance, this improvement cannot be directly related to the proposed pipelined TDMA, as differences in FFT implementations, domain configurations, and base  TDMA also affect the overall results. However, a comparison with other TDMA confirms that our GPU-based TDMA implementation is free from critical errors affecting performance, and the proposed method achieves competitive performance.

As shown in the scaling test, the pipelined strategy consistently enhances scalability by reducing non-scalable execution time, particularly  in scenarios involving large GPU counts or reduced domain sizes per GPU.
As the number of GPUs increases, the relative cost of non-scalable stages becomes more pronounced, and because of this, the hiding mechanism—especially for communication—of the pipelined algorithm remains effective, or even more so.
Given the same level of performance with \citet{sanhueza2025pencil} at 64 GPUs, and the fact that \citet{sanhueza2025pencil} verified the usefulness of a similar method without a pipelining algorithm up to 256 GPUs, the current implementation is expected to remain effective at least up to 256 GPUs, assuming similar conditions.

\section{Conclusion}

A highly scalable tridiagonal matrix algorithm (TDMA), \textbf{Pipelined-TDMA}, is developed by eliminating most of the non-scalable overhead in divide-and-conquer–based TDMA methods for multi-GPU systems.
The core contribution lies in the pipelined execution of multiple TDM systems, which hides both communication latency and the performance degradation caused by low GPU occupancy in intermediate TDM system (Stage~B).
This is achieved through computation–communication overlap and concurrent kernel execution.
As long as communication time remains shorter than computation time—a condition usually satisfied in practical scenarios—the pipelined structure effectively masks communication overhead.
The proposed pipelining strategy is integrated into a divide-and-conquer TDMA solver that extends the GPU-based PPT method of~\citet{ha2021multi} with the even-distribution scheme of~\citet{kim2021pascal_tdma}.

To further improve scalability, the batch size is optimized prior to the main computation by evaluating a wide range of candidate values.  
Larger batch sizes generate larger TDM systems, which improve computational performance.
However, excessively large batch sizes increase the impact of pipeline underutilization during its terminal phases, thereby exposing greater unhidden communication latency.
The proposed batch optimization algorithm efficiently balances these trade-offs, delivering robust performance across varying GPU counts and problem sizes.

Performance evaluations on up to 64 NVIDIA A100 GPUs demonstrate that the pipelined approach significantly reduces non-scalable execution time and enhances both strong and weak scalability.
In weak scaling tests with 1 billion grid cells per GPU, Pipelined-TDMA maintains near-ideal scalability.
In strong scaling tests on a 4 billion grid cells, it achieves 74.7\% parallel efficiency at 64 GPUs relative to the 4-GPU baseline.
Compared to the non-pipelined version on 64 GPUs, the non-scalable stages' execution time decreases from 14\% of scalable stages' execution time to 1.2\% in weak scaling test and from 74\% to 21\% in strong scaling test, excluding the pipeline’s initial and terminal phases.
Relative to the raw TDMA implementation of~\citet{ha2021multi}, the Pipelined-TDMA achieves a 16.2× speedup in strong scaling and a 3.3× speedup in weak scaling at 64 GPUs.

As an application, the proposed TDMA solver is integrated into an ADI-based fractional-step method with second-order spatial discretization for simulating incompressible flows. 
It is embedded within an FFT-based Poisson solver under a one-dimensional (1D) slab-type domain decomposition. 
The scalability limitation of the original flow solver by~\citet{ha2021multi}, which performs efficiently only on a small number of GPUs, is addressed through the enhanced scalability of the TDMA.
The 1D strategy offers several advantages for moderately large-scale motion : it avoids global all-to-all communication during FFT-based Poisson solves, simplifies implementation, and reduces post-processing costs—especially in canonical channel and boundary-layer flows, where spanwise and streamwise Fourier transform is commonly used.

Two-dimensional (2D) domain decomposition is essential for extremely large-scale simulations, and the proposed TDMA is expected to remain a critical component under such configurations. 
A pencil-type decomposition, which combines divide-and-conquer methods along one direction with global transposes along the other, is well-suited for solving the Poisson equation. 
In this configuration, since the TDMA direction requires significantly finer partitioning than the FFT direction to enable intra-node all-to-all communication, the scalability of TDMA will remain a key concern.
Furthermore, increasing the number of partitions along the TDMA direction shifts the communication burden from FFT to TDMA. 
Given that TDMA generally incurs a higher computational cost than FFT, this shift enables communication latency to be more effectively hidden during TDMA execution.
Apart from the Poisson solver, the momentum equation involves only the solution of TDM systems and does not require FFTs, which incur all-to-all communication. As a result, the applicability of the proposed TDMA remains unaffected by the choice between 1D and 2D decomposition.

Although Pipelined-TDMA has demonstrated its utility in canonical turbulent boundary layer and channel flow simulations, its applicability extends far beyond this context.
TDMA are fundamental to implicit time integration and FFT-based Poisson solvers, and are widely used in large-scale simulation applications such as urban large-eddy simulations~\citep{yang2023multi}, weather prediction models (e.g., WRF~\citep{skamarock2008description}), and other simulations on Cartesian grids employing second-order accurate spatial discretization schemes.
TDMA also supports high-order spatial accuracy schemes, which can be integrated into applications such as flow solvers~\citep{you2006high} and wave propagation models~\citep{jiang2023explicit} for high-fidelity simulations.

\section*{CRediT authorship contribution statement}
\textbf{Seungchan Kim}: Idea formulation, Methodology, Software, Writing original draft, Investigation, Validation. \textbf{Jihoo Kim}: Investigation, Idea formulation. \textbf{Sang{hyun} Ha}: Writing original draft, Software. \textbf{Donghyun You}: Conceptualization, Funding acquisition, Writing original draft.

\section*{Declaration of competing interest}
The authors declare that they have no known competing financial interests or personal relationships that could have appeared to influence the work reported in this paper.

\section*{Acknowledgements}
This work was supported by the National Research Foundation of Korea (NRF) under grant number RS-2023-00282764. The authors gratefully acknowledge {Mario Rüttgers for providing} computing time on the supercomputer JURECA-DC~\cite{JURECA} at Forschungszentrum Jülich, provided under project grant no. jhpc54.

\bibliographystyle{unsrt}




\end{document}